\documentclass[apj]{emulateapj}
\usepackage{color}

\slugcomment{Accepted version October 13, 2011}
\shorttitle{}
\shortauthors{FUKUI ET AL.}

\begin{document}
\title{A Detailed Study of the Molecular and Atomic Gas Toward the\\ $\gamma$-ray SNR RX J1713.7$-$3946: Spatial TeV $\gamma$-ray and ISM Gas Correspondence}

\author{Y. Fukui\altaffilmark{1}, H. Sano\altaffilmark{1}, J. Sato\altaffilmark{1}, K. Torii\altaffilmark{1}, H. Horachi\altaffilmark{1}, T. Hayakawa\altaffilmark{1}, N. M. McClure-Griffiths\altaffilmark{2}, G. Rowell\altaffilmark{3}, T. Inoue\altaffilmark{4}, S. Inutsuka\altaffilmark{1}, A. Kawamura\altaffilmark{1,5}, H. Yamamoto\altaffilmark{1}, T. Okuda\altaffilmark{1}, N. Mizuno\altaffilmark{1,5}, T. Onishi\altaffilmark{1,6}, A. Mizuno\altaffilmark{7} and H. Ogawa\altaffilmark{6}}

\affil{$^1$Department of Physics and Astrophysics, Nagoya University, Furo-cho, Chikusa-ku, Nagoya, Aichi 464-8601, Japan; fukui@a.phys.nagoya-u.ac.jp}
\affil{$^2$CSIRO Astronomy and Space Science, PO Box 76, Epping NSW 1710, Australia}
\affil{$^3$School of Chemistry and Physics, University of Adelaide, Adelaide 5005, Australia}
\affil{$^4$Department of Physics and Mathematics, Aoyama Gakuin University, Fuchinobe, Chuou-ku, Sagamihara, Kanagawa 252-5258, Japan}
\affil{$^5$National Astronomical Observatory of Japan, Mitaka, Tokyo 181-8588, Japan}
\affil{$^6$Department of Astrophysics, Graduate School of Science, Osaka Prefecture University, 1-1 Gakuen-cho, Naka-ku, Sakai, Osaka 599-8531, Japan}
\affil{$^7$Solar-Terrestrial Environment Laboratory, Nagoya University, Furo-cho, Chikusa-ku, Nagoya, Aichi 464-8601, Japan}

\begin{abstract}
RX J1713.7$-$3946 is the most remarkable TeV $\gamma$-ray SNR which emits $\gamma$-rays in the highest energy range. We made a new combined analysis of CO and \ion{H}{1} in the SNR and derived the total protons in the interstellar medium (ISM). We have found that the inclusion of the \ion{H}{1} gas provides a significantly better spatial match between the TeV $\gamma$-rays and ISM protons than the H$_2$ gas alone. In particular, the southeastern rim of the $\gamma$-ray shell has a counterpart only in the \ion{H}{1}. The finding shows that the ISM proton distribution is consistent with the hadronic scenario that comic ray (CR) protons react with ISM protons to produce the $\gamma$-rays. This provides another step forward for the hadronic origin of the $\gamma$-rays by offering one of the necessary conditions missing in the previous hadronic interpretations. We argue that the highly inhomogeneous distribution of the ISM protons is crucial in the origin of the $\gamma$-rays. Most of the neutral gas was likely swept up by the stellar wind of an OB star prior to the SNe to form a low-density cavity and a swept-up dense wall. The cavity explains the low-density site where the diffusive shock acceleration of charged particles takes place with suppressed thermal X-rays, whereas the CR protons can reach the target protons in the wall to produce the $\gamma$-rays. The present finding allows us to estimate the total CR proton energy to be $\sim $10$^{48}$ ergs, 0.1 \% of the total energy of a SNe.
\end{abstract}

\keywords{cosmic rays -- gamma rays: ISM  -- ISM: individual objects (RX J1713.7--3946) -- ISM: clouds -- ISM: molecules -- ISM: \ion{H}{1}}

\section{Introduction}
It is a long-standing question how the cosmic ray (CR) protons, the major constituent of cosmic rays, are accelerated in the interstellar space. Supernova remnants (SNRs) are the most likely candidate for the acceleration because the high-speed shock waves offer an ideal site for diffusive shock acceleration (DSA) \citep[e.g.,][]{bell1978, blandford1978}. The principal site of CR proton acceleration is, however, not yet identified observationally in spite of a number of efforts to address this issue. 

RX J1713.7$-$3946 is the brightest and most energetic TeV $\gamma$-ray SNR detected in the Galactic plane survey with H.E.S.S. \citep{aharonian2006a}  and is a primary candidate where the origin of the $\gamma$-rays may be established. Discovery of the SNR was made in X-rays with {\it ROSAT} \citep{pfeffermann1996} and {\it ASCA} showed that the X-rays are non-thermal synchrotron emission with no thermal features \citep{koyama1997}. TeV $\gamma$-rays were first detected by CANGAROO \citep{enomoto2002} and, subsequently, H.E.S.S. resolved the shell-like TeV $\gamma$-ray distribution with a $\sim 0\fdg 1$ degree point spread function \citep{aharonian2004, aharonian2006b, aharonian2007}. The $\gamma$-rays are emitted via two mechanisms, either leptonic or hadronic, closely connected to the CR particles and it is important to understand which mechanism is working in the SNR. The leptonic process includes the inverse Compton effect of CR electrons which energize the low energy photons of the cosmic microwave background and additional soft photon fields (e.g., infrared). The hadronic process includes the neutral pion decay into $\gamma$-rays following the reaction of CR protons with the low energy target protons in the interstellar medium (ISM). Considerable work has been devoted to explaining the $\gamma$-ray emission in the framework of leptonic and hadronic scenarios \citep{aharonian2006b, porter2006, katz2008, berezhko2008, ellison2008, tanaka2008, morlino2009, acero2009, ellison2010, patnaude2010, zirakashvili2010, abdo2011, fang2011}. 

The molecular gas interacting with the SNR was discovered in $V_{\mathrm{LSR}}$, the velocity with respect to the local standard of rest, around $-7$ km s$^{-1}$ at 2.6 arcmin resolution based on NANTEN Galactic plane CO survey and the distance of the SNR was determined to be 1 kpc by using the flat rotation curve of the Galaxy \citep{fukui2003}. This determination offered a robust verification for a small distance of 1 kpc and revised an old value 6 kpc derived from lower resolution CO observations \citep{slane1999}. Studies of X-ray absorption suggested a similar distance 1 kpc under an assumption of uniform foreground gas distribution \citep{koyama1997}, but the local bubble of \ion{H}{1} located by chance toward the SNR makes the X-ray absorption uncertain in estimating the distance \citep{slane1999, matsunaga2001}. A subsequent careful analysis of the X-ray absorption favors the smaller distance \citep{cassamchenai2004}. At 1 kpc the SNR has a radius of 9 pc and an age of 1600 yrs \citep{fukui2003, wang1997} and the expanding shock front has a speed of 3000 km s$^{-1}$ \citep{zirakashvili2007, uchiyama2003, uchiyama2007}. 
\citet{moriguchi2005}  showed further details of the NANTEN CO distribution and confirmed the identification of the interacting molecular gas at $-20$--0 km s$^{-1}$ by \citet{fukui2003}. Most recently, \citet{sano2010}  showed that the SNR harbors the star forming dense cloud core named peak C at similar $V_{\mathrm{LSR}}$ and argued that X-rays are bright around the core, reinforcing the association of the molecular gas. 

Importantly, the associated molecular gas with the SNR opened a possibility to identify target protons where the hadronic process is working. If the cosmic ray density is nearly uniform, we expect the $\gamma$-ray distribution mimics that of the interstellar target protons. Some nearby molecular clouds show good spatial correlation with relatively high resolution $\gamma$-ray images, clearly verifying that the hadronic process is working to produce $\gamma$-rays in the cosmic ray sea (e.g., \citeauthor{bertsch1993} \citeyear{bertsch1993} and the references therein.). A detailed comparison between the ISM protons and the recent high-resolution $\gamma$-ray images of H.E.S.S. sources is useful to test the correlation, although such a test was not possible until recently in the preceding low resolution $\gamma$-ray observations at degree-scale resolution. 
\citet{aharonian2006b} compared the NANTEN CO distribution with the H.E.S.S. TeV $\gamma$-ray image and examined both leptonic and hadronic scenarios as the origin of the $\gamma$-rays. By adopting annular averaging of TeV $\gamma$-rays and CO in the shell \citep[see Figure 17 of][]{aharonian2006b}, these authors found that TeV $\gamma$-rays are fairly well correlated with CO, whereas the correlation is not complete in the sense that the southeastern rim of the TeV $\gamma$-ray shell has no counterpart in CO. The complete identification of target ISM protons thus remained unsettled in the hadronic scenario. 

The $\gamma$-rays and X-rays are significantly enhanced toward the clumpy molecular gas at a pc scale as first shown by \citet{fukui2003}. 
A strong connection among the molecular gas, the $\gamma$-rays, the X-rays, and perhaps cosmic rays is therefore suggested \citep{fukui2003, fukui2008, sano2010, zirakashvili2010}. Most of the previous models of the $\gamma$- and X-rays cited earlier assume more or less uniform density distribution of the ISM in the SNR, whereas the observations of the ISM indicate that the actual distribution is highly inhomogeneous, with density varying by a factor 100 or more around the SNR. In addition, Galactic-scale studies of $\gamma$-rays suggest that there is "dark gas" which is not detectable in CO or in \ion{H}{1} but still contributes to the $\gamma$-rays and visual extinction \citep{grenier2005, ade2011}. Such gas may be either cold \ion{H}{1} or H$_2$ with no detectable CO. So, it is important to consider \ion{H}{1} carefully in order to have a comprehensive understanding of the ISM protons. In the present paper, we shall use the term "dark \ion{H}{1}" for observed \ion{H}{1} with significantly lower brightness than the surroundings. So, "dark \ion{H}{1}" does not necessarily mean "dark gas" above, while they might be linked.

We here present a combined analysis of both the $^{12}$CO($J$=1--0) and \ion{H}{1} datasets in order to clarify the distribution of the ISM protons in RX J1713.7$-$3946. The quantitative analysis is made mainly by using the $^{12}$CO($J$=1--0) and \ion{H}{1} data here and a comparative study of the $^{12}$CO($J$=1--0, 2--1) transitions is a subject of a forthcoming paper. A theoretical work of magneto-hydrodynamics which incorporates fully the inhomogeneous ISM in RX J1713.7$-$3946 is complimentary to the present observational paper and will be published separately (Inoue, Yamazaki, Inutsuka $\&$ Fukui 2011; hereafter IYIF2011). The present paper is organized as follows. 
Section \ref{section:datasets} gives the description of the CO and \ion{H}{1} datasets. 
Section \ref{section:analysis} consists of five Sub-sections; Sub-section 3.1 gives a combined data set of CO and \ion{H}{1}, Sub-sections 3.2--3.4 the distribution of total ISM protons in the SNR with an emphasis on dark \ion{H}{1}, the cold and dense atomic phase of ISM protons, and Sub-section 3.5 comparison with the $\gamma$-rays. 
Section \ref{section:discussion} consists of two Sub-sections; Sub-section 4.1 describes the initial distribution of the ISM before the stellar explosion, and Sub-section 4.2 the scheme of particle acceleration and its relationship with the $\gamma$-rays. The conclusions are given in Section \ref{section:conclusions}.

\begin{figure*}
\begin{center}
\includegraphics[width=179mm,clip]{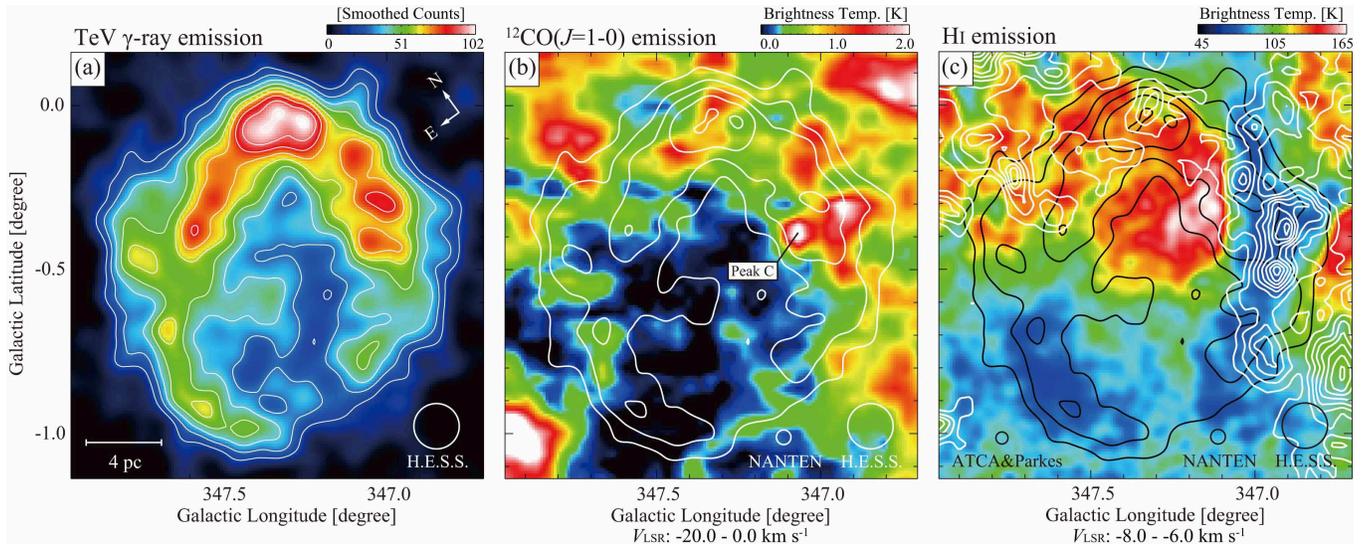}
\caption{(a) The H.E.S.S. TeV $\gamma$-ray distribution of RX J1713.7$-$3946 in smoothed excess counts above the cosmic-ray background (see Figure 2 of \citeauthor{aharonian2007} \citeyear{aharonian2007}). Contours are plotted every 10 smoothed counts from 20 smoothed counts. (b) Averaged brightness temperature distribution of $^{12}$CO($J$=1--0) emission in a velocity range of $V_{\mathrm{LSR}}$ = $-20$ km s$^{-1}$ to $0$ km s$^{-1}$ is shown in color \citep{fukui2003, moriguchi2005}. White contours show the H.E.S.S. TeV $\gamma$-ray distribution and are plotted every 20 smoothed counts from 20 smoothed counts. (c) Averaged brightness temperature distribution of \ion{H}{1} emission obtained by ATCA and Parkes in a velocity range from $V_{\mathrm{LSR}}$ = $-8$ km s$^{-1}$ to $-6$ km s$^{-1}$ \citep{mccluregriffiths2005} is shown in color. White contours show the $^{12}$CO($J$=1--0) brightness temperature integrated in the same velocity range every 1.0 K km s$^{-1}$ ($\sim$3$\sigma$).}
\end{center}
\end{figure*}%

\section{Datasets of CO, \ion{H}{1} and TeV $\gamma$-rays}\label{section:datasets}

The $^{12}$CO($J$=1--0) data at 2.6 mm wavelength were taken with NANTEN 4m telescope in 2003 April and are identical with those published by \citet{moriguchi2005}. The system temperature of the SIS receiver was $\sim $250 K in the single side band including the atmosphere toward the zenith. The beam size of the telescope was $2\farcm 6$ at 115 GHz and we adopted a grid spacing of $2.0\arcmin$ in these observations. We adopt hereafter the brightness temperature (K) as the spectral line intensity scale. The velocity resolution and rms noise fluctuations are 0.65 km s$^{-1}$ and 0.3 K, respectively.

The $^{12}$CO($J$=2--1) data at 1.3 mm wavelength were taken with NANTEN2 4m telescope in the period from August to November in 2008 and part of the dataset was published by \citet{sano2010}. The frontend was a 4 K cooled Nb SIS mixer receiver and the single-side-band (SSB) system temperature was $\sim $250 K, including the atmosphere toward the zenith.  The telescope had a beam size of $90\arcsec$ at 230 GHz. We used an acoustic optical spectrometers (AOS) with 2048 channels having a bandwidth of 390 km s$^{-1}$ and resolution per channel of 0.38 km s$^{-1}$. Observations in $^{12}$CO($J$=2--1) were carried out in the on-the-fly (OTF) mode, scanning with an integration time of 1.0 to 2.0 sec per point. The chopper wheel method was employed for the intensity calibration and the rms noise fluctuations were better than 0.66 and 0.51 K per channel with 1.0 and 2.0 sec integrations, respectively. An area of 2.25 square degrees in a region of 346.7 deg $\leq $ $l$ $\leq $ 348.2 deg and $-1.3$ deg $\leq $ $b$ $\leq $ 0.4 deg was observed.

The \ion{H}{1} data at 21 cm wavelength are from the Southern Galactic Plane Survey \citep[SGPS;][]{mccluregriffiths2005} and combined from the Australia Telescope Compact Array and the Parkes Radio Telescope. The beam size of the dataset was 2.2 arcmin and we adopted a grid spacing of $40\arcsec$ toward the RX J1713.7$-$3946 in the current analysis. The velocity resolution and typical rms noise fluctuations were 0.82 km s$^{-1}$ and 1.9 K, respectively.

\cite{moriguchi2005} showed an analysis of the $^{12}$CO($J$=1--0) distribution over 100 km s$^{-1}$ with a coarse velocity window of 10 km s$^{-1}$ in order to test association with the SNR. These authors showed that the velocity range $V_{\mathrm{LSR}}$ = $-20$ to 0 km s$^{-1}$ has convincing signs of association with the SNR. We adopt in the present work the velocity interval, $-20$--0 km s$^{-1}$, for the associated ISM and present detailed $^{12}$CO ($J$=1--0 and 2--1) and \ion{H}{1} data every 1 km s$^{-1}$ (see Figure A in Appendix A). 

For the H.E.S.S. $\gamma $-ray data we used the combined H.E.S.S. image shown in Figure 2 of \citet{aharonian2007}. Data of 2004 and 2005 are used for this smoothed, acceptance-corrected gamma-ray excess image. The TeV image utilizes minimum 3 H.E.S.S. telescopes in event reconstruction to obtain a Gaussian standard deviation of $0\fdg 06$ or FWHM of $0\fdg 14$ (8.3$\arcmin$).

\begin{figure*}
\begin{center}
\includegraphics[width=179mm,clip]{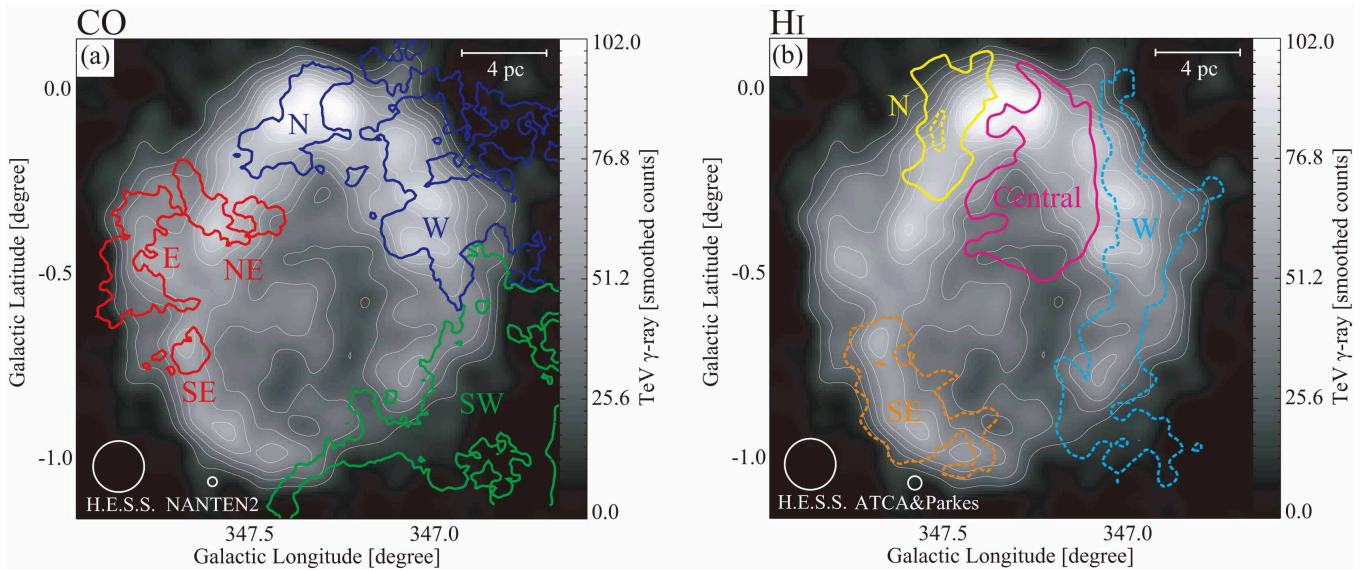}
\caption{(a) Schematic of the identified $^{12}$CO($J$=2--1) clouds is shown in colored contours. The image and white contours show the TeV $\gamma$-ray distribution (Figure 1(a)). The integration velocity ranges are follows; $-8$--$-2$ km s$^{-1}$ (contour level: 2.3 K km s$^{-1}$) for the SW cloud (CO), $-17$--$-5$ km s$^{-1}$ (contour level: 4.9 K km s$^{-1}$) for the W and N clouds (CO) and $-6$--$0$ km s$^{-1}$ (contour level: 2.7 K km s$^{-1}$) for the NE, E and SE clouds (CO). (b) The locations of the identified \ion{H}{1} clouds are shown in colored contours. The gray scale image and white contours show the TeV $\gamma$-ray distribution. The solid contours are for \ion{H}{1} emission and the dashed contours are for dark \ion{H}{1}. The integration velocity ranges are as follows; $-8$--$-2$ km s$^{-1}$ (contour level: 214 K km s$^{-1}$) for the W cloud (\ion{H}{1}), $-12$--$-4$ km s$^{-1}$ (contour level: 780 K km s$^{-1}$) for the central and SE clouds (\ion{H}{1}), and $-14$--$-11$ km s$^{-1}$ (contour level: 399 K km s$^{-1}$) for the N cloud (\ion{H}{1}).}
\end{center}
\end{figure*}%

\section{Combined analysis of the CO and \ion{H}{1} data}\label{section:analysis}

\subsection{Distribution of CO and \ion{H}{1}}

Figure 1(a) shows TeV $\gamma$-ray distribution toward RX J1713.7$-$3946 obtained by H.E.S.S. and Figure 1(b)  shows a velocity averaged distribution of $^{12}$CO($J$=1--0) overlayed on the TeV $\gamma$-ray distribution. The $^{12}$CO($J$=1--0) intensity becomes larger in the north to the Galactic plane than in the south and the most prominent features above 0.7 K are located in the northwest. The general $^{12}$CO($J$=1--0) distribution is shell-like associated with the $\gamma$-ray shell, showing weaker or no CO emission in part of the south. There are two regions where $^{12}$CO($J$=1--0) delineates particularly well the outer boundary of the shell in the southwest and east. In addition, we see some of the $^{12}$CO($J$=1--0) features are located within the shell including the prominent peak C at ($l$, $b$) = (347$\fdg$07, -0$\fdg$40). 

The $^{12}$CO($J$=2--1) distribution is qualitatively similar to the $^{12}$CO($J$=1--0) distribution. A typical ratio of the [$J$=1-0]/[$J$=2-1] line intensities is $\sim$0.6, consistent with what are derived in the other molecular clouds without heat source \citep[e.g.][]{ohama2010, torii2011}. We tentatively choose from Figure A three major CO clouds, W, N and SW, and three minor ones, E, NE and SE, for the sake of discussion as schematically shown in Figure 2(a), where we use $^{12}$CO($J$=2--1) data by taking an advantage of higher angular resolution. 

Figure 1(c) shows an overlay of the \ion{H}{1} distribution superposed on the $^{12}$CO($J$=1--0) intensity in a velocity range of $-8.0$--$-6.0$ km s$^{-1}$. 
The average \ion{H}{1} brightness temperature ranges from 60 to 150 K and becomes higher toward the Galactic plane. 
The brightest \ion{H}{1} of $\sim $150 K, the central cloud, is located toward the center of the SNR [($l$, $b$)=(347$\fdg$25, $-0\fdg$38)] where little $^{12}$CO($J$=1--0) is seen. 
We find dark \ion{H}{1} clouds of around 60 K in the west (W cloud) and in the southeast (SE cloud). 
These dark \ion{H}{1} clouds are not due to absorption of the radio continuum radiation which is very weak toward the SNR \citep{lazendic2004}. 
The dark \ion{H}{1} W cloud well corresponds to the $^{12}$CO($J$=1--0) distribution, showing sharp edges both toward the east and west. The dark \ion{H}{1} SE cloud has almost no counterpart in CO. The relatively bright \ion{H}{1} emission is seen in the north of the SNR (N cloud). The N cloud tends to be located toward $^{12}$CO($J$=1--0) peaks, whereas the \ion{H}{1} brightness shows a non-monotonic, more complicated behavior than in the W cloud. In the northeast, we find a rim of relatively lower \ion{H}{1} brightness of $\sim $100 K toward {($l$, $b$)}=(347$\fdg$5, $-0\fdg$25) that lies along the $\gamma $-ray shell. A schematic of the four main \ion{H}{1} clouds is given in Figure 2(b). The good correspondence of the \ion{H}{1} clouds with the CO and the $\gamma$-rays supports that the \ion{H}{1} is physically associated with the SNR.

In Figure 3 we show typical \ion{H}{1} and CO profiles in the four main \ion{H}{1} clouds. Figure 3 indicates that the \ion{H}{1} emission is generally peaked at $-10$ km s$^{-1}$ with small hints of saturation, confirming that the \ion{H}{1} is associated with the SNR and is generally not optically thick. We find that narrow \ion{H}{1} dips having depths of 20--30 K often correspond to $^{12}$CO($J$=1--0) emission features in the N and W clouds. The linewidths of the narrow \ion{H}{1} dips are as small as a few km s$^{-1}$. It is likely that these \ion{H}{1} dips represent residual \ion{H}{1} in cold CO gas seen as self-absorption. The broad \ion{H}{1} dip in the SE cloud is also ascribed to self-absorption as argued into detail in Sub-section 3.3. We show \ion{H}{1} expected profiles of the background \ion{H}{1} emission with a straight-line approximation as dashed areas in Figure 3 \citep[e.g.,][]{sato1978}.

\begin{figure*}
\begin{center}
\includegraphics[width=179mm,clip]{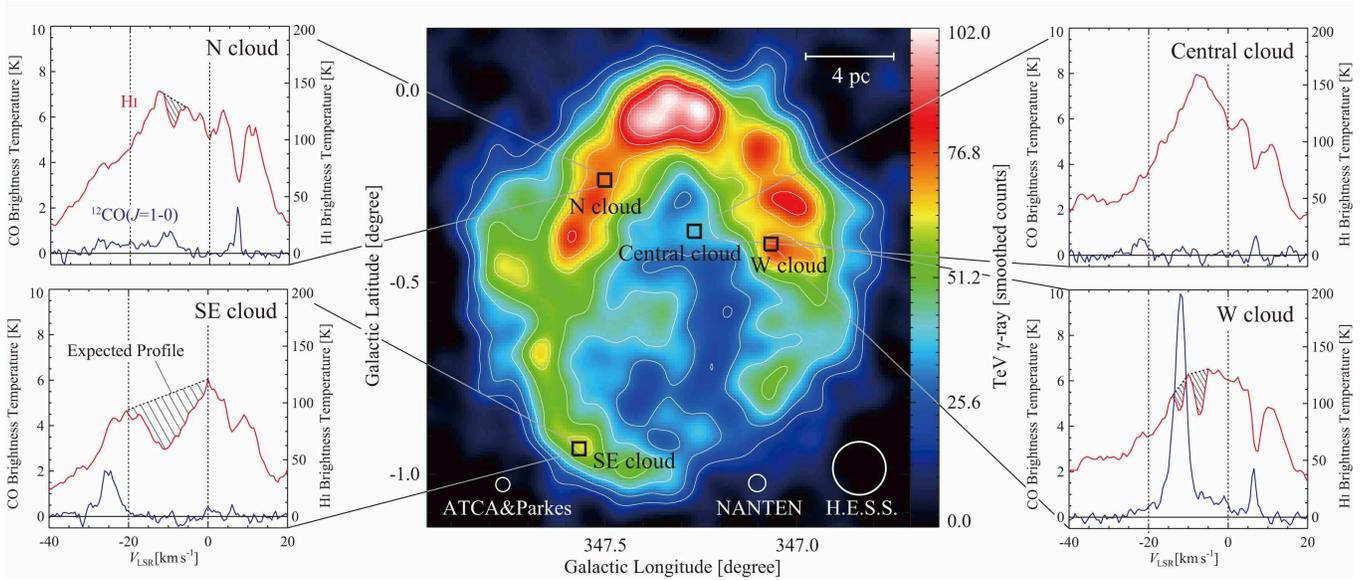}
\caption{$^{12}$CO($J$=1--0) and \ion{H}{1} profiles {at} the four \ion{H}{1} clouds; the N cloud ($l$, $b$) = (347$\fdg$50, $-$0$\fdg$23), the SE cloud ($l$, $b$) = (347$\fdg$57, $-$0$\fdg$93), the central cloud ($l$, $b$) = (347$\fdg$27, $-$0$\fdg$37), and the W cloud ($l$, $b$) = (347$\fdg$07, $-$0$\fdg$40). The positions are denoted in the H.E.S.S. TeV $\gamma$-ray distribution. The shaded area shows expected profiles behind the self-absorption.}
\end{center}
\end{figure*}%

\begin{figure*}
\begin{center}
\includegraphics[width=179mm,clip]{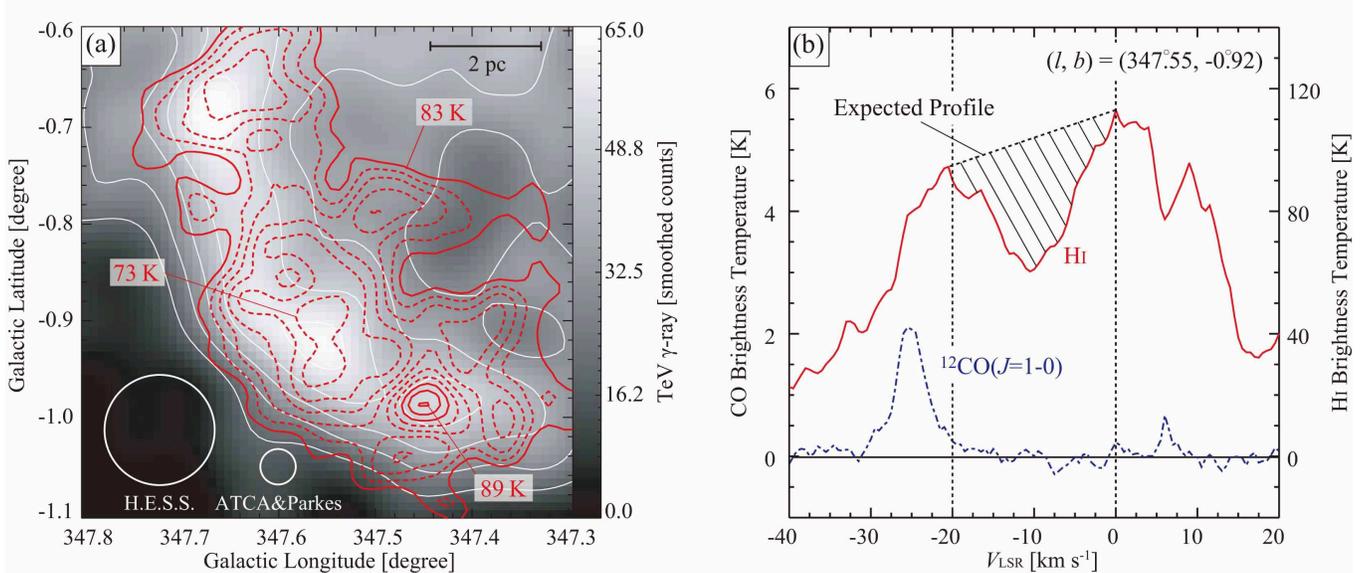}
\caption{(a)The H.E.S.S. TeV $\gamma$-ray distribution toward the SE cloud \citep{aharonian2007}. Red contours show averaged \ion{H}{1} brightness temperature distribution in a velocity range from $-$15 km s$^{-1}$ to $-$5 km s$^{-1}$ \citep{mccluregriffiths2005}. (b) The \ion{H}{1} and $^{12}$CO($J$=1--0) spectra at ($l$, $b$) =(347$\fdg$55, $-$0$\fdg$92). The shaded area shows an expected \ion{H}{1} profile.}
\end{center}
\end{figure*}%

\subsection{Molecular protons}

In order to covert the $^{12}$CO($J$=1--0) intensity into the total molecular column density, we use an X factor, which is defined as X (cm$^{-2}$/K km s$^{-1}$) = $N$(H$_2$) (cm$^{-2}$)/W($^{12}$CO) (K km s$^{-1}$). In order to derive an X factor, the $^{12}$CO($J$=1--0) intensity is compared with the cloud dynamical mass (virial mass), or with the $\gamma$-rays produced via interaction of cosmic ray protons with molecular clouds. An X factor therefore accounts for the total hadronic mass and is observationally uniform in the Galactic disk \citep[e.g.][]{fukui2010}. We here adopt an X factor of 2.0$\times $10$^{20}$ $W$($^{12}$CO) (cm$^{-2}$/K km s$^{-1}$) derived from the $\gamma$-rays and $^{12}$CO($J$=1--0) intensity in the Galaxy \citep{bertsch1993}. We double the H$_2$ column density to derive the ISM protons in molecular form as shown in Figure 7(a). Compared with the $^{12}$CO($J$=1--0) line, the $^{12}$CO ($J$=2--1) line is not a common probe of the molecular mass. This is in part because the $^{12}$CO($J$=2--1) emission samples a smaller portion, having a higher excitation condition, of a molecular cloud than traced by the $^{12}$CO($J$=1--0) emission. We estimate for instance that a typical fraction in area of the $^{12}$CO($J$=2--1) emission to the $^{12}$CO($J$=1--0) emission is about 70--80 $\%$ at the half-intensity level convolved to the same beam size in the present region from the CO data in Figure A. 

\subsection{Atomic protons}
\subsubsection{Optically thin case}

We use the 21cm \ion{H}{1} transition to estimate the atomic proton column density. A usual assumption is that the \ion{H}{1} emission is optically thin and the following relationship is used to calculate the \ion{H}{1} column density;
{\begin{eqnarray}
N_{\mathrm{p}}\mathrm{(\mbox{\ion{H}{1}})} = 1.823 \times 10^{18} \int T_L(V) dV \:\: \mathrm{(cm^{-2})}
\label{equation:optthin}
\end{eqnarray}
where $T_L(V)$ is the observed \ion{H}{1} brightness temperature (K) \citep{dickey1990}. We note that this simple assumption is usually valid and apply equation (\ref{equation:optthin}) to the regions where no \ion{H}{1} dips are seen. It is certain that the narrow \ion{H}{1} dips in the W and N clouds represent self-absorption by cold residual \ion{H}{1} in CO gas from their exact coincidence with CO in velocity. The most prominent dark \ion{H}{1} cloud, the SE cloud, shows large linewidths, not so common as self-absorption. We shall examine if the SE cloud represents self-absorption in the followings. 

\subsubsection{The dark \ion{H}{1} SE cloud }

\begin{figure*}
\begin{center}
\includegraphics[width=179mm,clip]{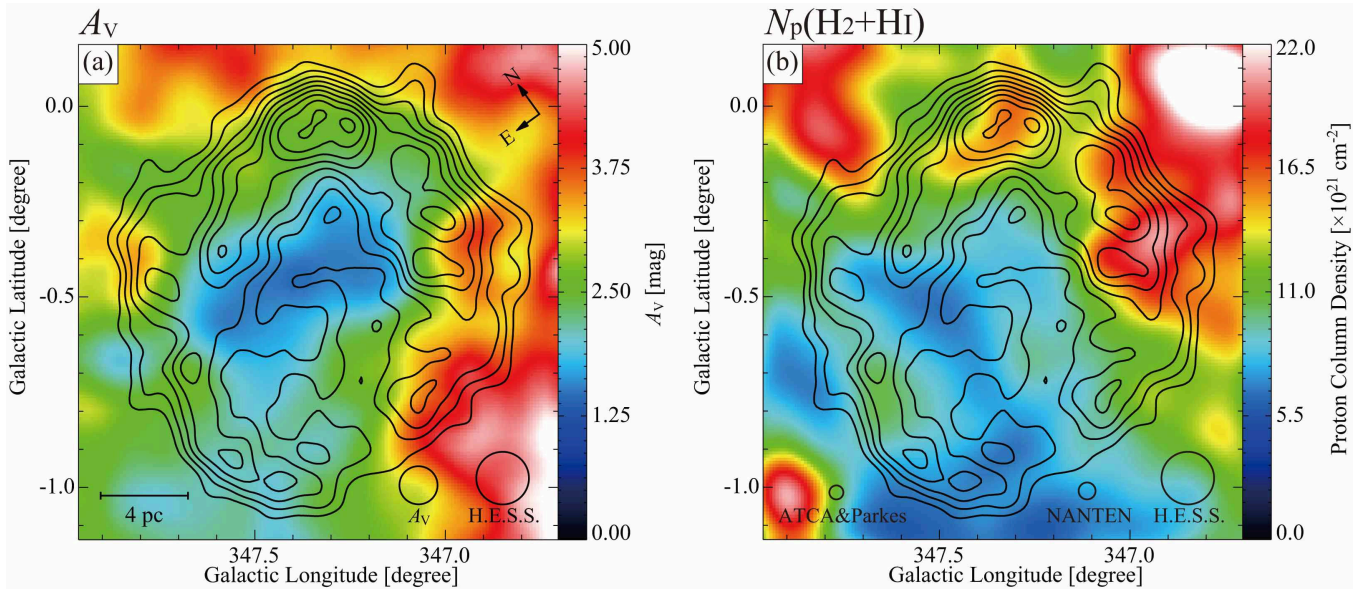}
\caption{(a)$A_{\mathrm{V}}$ distribution \citep{dobashi2005} is shown in color. Contours are the same as in Figure 1(a). (b) Distribution of column density of the total ISM protons estimated from both CO and \ion{H}{1} in a velocity range from $V_{\mathrm{LSR}}$ = $-20$ to 10 km s$^{-1}$. Here the \ion{H}{1} self-absorption is taken into account. Contours are the same as in Figure 1(a).}
\end{center}
\end{figure*}%

We first show the integrated intensity image of the SE cloud in Figure 4(a). The \ion{H}{1} contours are every 3.9 $\sigma $ noise level and shows significant details not apparent in Figure 1(c), where a coarser color code is used. We find the \ion{H}{1} brightness variation is generally well correlated with the shell of $\gamma$-rays in gray scale in the sense that \ion{H}{1} brightness decreases toward the enhanced $\gamma$-rays. This trend lends a support for physical connection of the SE cloud with the $\gamma$-ray shell and may be interpreted as due to decrease in spin temperature with density increase in the self-absorbing \ion{H}{1} gas (see Sub-section 3.3.3). No $^{12}$CO($J$=1--0) emission is seen toward the SE cloud, expect for a possible small counterpart at ($l$, $b$) = (347$\fdg$64, -0$\fdg$72) and $V_{\mathrm{LSR}}$ = {$-6$--0 km s$^{-1}$} (Figure 2(a) and Figure A), suggesting that density of the SE cloud is lower than the CO clouds.

Figure 4(b) shows a typical \ion{H}{1} profile in the SE cloud having a deep and broad dip. The large velocity span of 20 km s$^{-1}$ is not so common as a self-absorption feature; in nearby dark clouds \ion{H}{1} self-absorption is generally narrow with a few km s$^{-1}$ in linewidth \citep[e.g.][]{krco2010}, whereas \ion{H}{1} self-absorption as broad as 10 km s$^{-1}$ is seen in giant molecular clouds \citep[e.g.][]{sato1978}. The SE cloud delineates the $\gamma$-ray shell (Figure 4(a)) and is possibly compressed gas by the wind of a high-mass star, the SN projenitor. We have investigated the velocity distribution of the SE cloud as given in Appendix A. We find that the SE cloud shows a strong velocity gradient which matches the blue-shifted part of an expanding swept-up shell. Such a shell is a natural outcome of the stellar-wind compression by the SN projenitor, supporting that the broad \ion{H}{1} dip is ascribed to the acceleration of \ion{H}{1} gas by the wind. A \ion{H}{1} stellar-wind shell in Pegasus driven by an early B star indeed shows a linewidth as large as 15 km s$^{-1}$ (Sub-section 4.1, \citeauthor{yamamoto2006} \citeyear{yamamoto2006}), similar to that of the SE cloud. The difference from the narrow \ion{H}{1} dips in the W and N clouds may be due to density; the SE cloud has lower density and is subject to stronger acceleration than the CO clouds with narrow \ion{H}{1} dips (see for further discussion Sub-section 4.1), whereas the CO clouds having higher density are less accelerated by the wind, making a systematic velocity gradient less clear in CO than in \ion{H}{1} (see Figure B3).

Figure 5(a) shows the distribution of the extinction $A_{\rm V}$ toward RX J1713.7$-$3946 \citep{dobashi2005}, and indicates that the SE cloud, as well as the rest of the shell, is traced by the enhanced optical extinction. This lends another support for the self-absorption interpretation of the SE cloud. Figure 5(b) shows the total (molecular and atomic) ISM proton column density both in the SNR (derived later in Sub-section 3.4) and in the foreground within 1 kpc, which is supposed to correspond mainly to the optical extinction. The total proton column density $N_{\mathrm{p}}$ of $\sim 10^{22}$ cm$^{-2}$ in Figure 5(b) corresponds to extinction of $\sim$4 magnitude if we adopt the relationship $N_{\mathrm{p}}$(cm$^{-2}$) = 2.5$\times$10$^{21}$ $\cdot $ $A_{\rm V}$ (magnitude) \citep{jenkins1974}. The extinction toward the SE cloud is 2--3 magnitude in Figure 5(a) and is consistent with the \ion{H}{1} self-absorption by considering the contamination by the foreground stars which tends to reduce $A_{\rm V}$ toward the Galactic plane.

In summary, we find it a reasonable interpretation that the SE cloud represents \ion{H}{1} self-absorption associated with the SNR shell. 

\subsubsection{Analysis of the \ion{H}{1} self-absorption dips}

\begin{figure*}
\begin{center}
\includegraphics[width=179mm,clip]{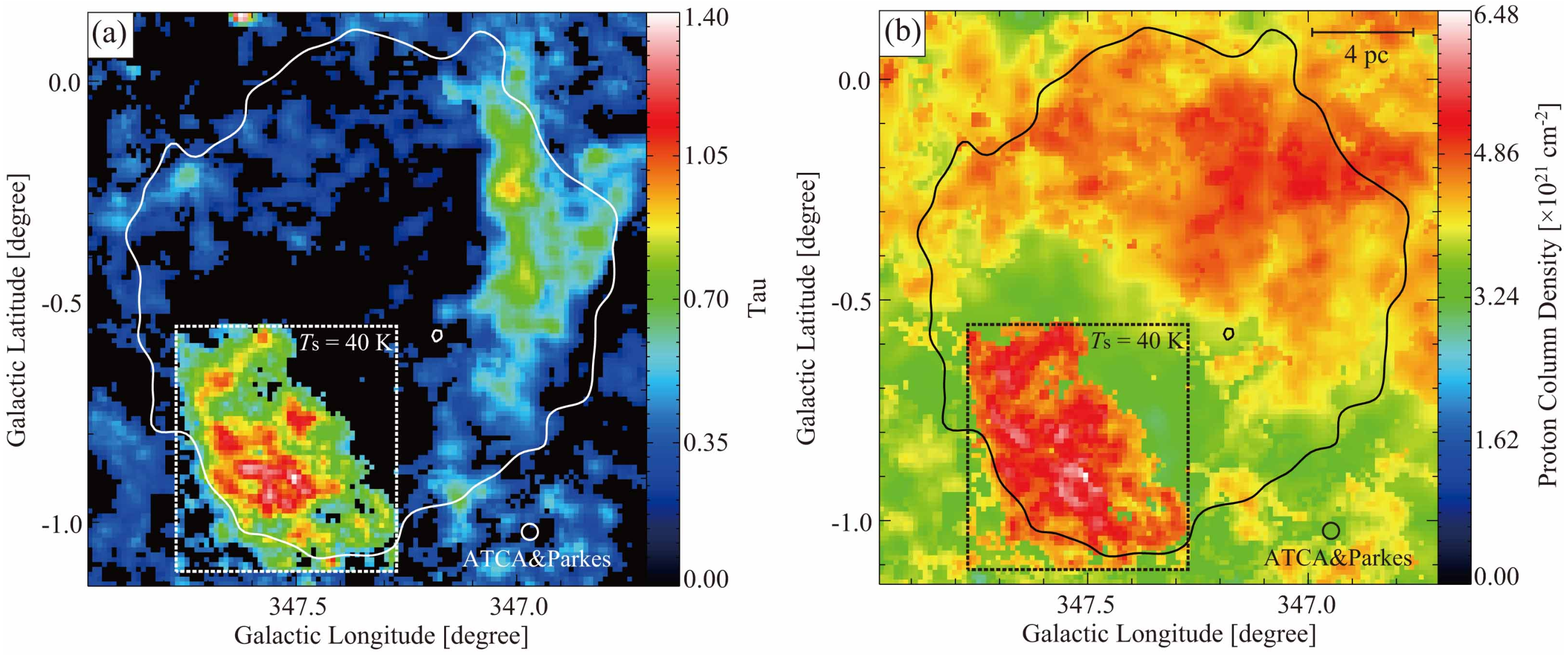}
\caption{(a) Distribution of peak optical depth of the \ion{H}{1} self-absorption. (b) Distribution of atomic proton column density, $N_{\mathrm{p}}$(\ion{H}{1}), estimated for the \ion{H}{1} emission and self-absorption. The velocity range in the both figures is from $-$20 km s$^{-1}$ to 0 km s$^{-1}$. Contours show the H.E.S.S. TeV $\gamma$-rays distribution \citep{aharonian2007} and are plotted at 20 smoothed counts. We assume spin temperatures $T_{\rm s}$ of 40 K and 10 K, inside and outside the dotted box toward the SE cloud, respectively.}
\end{center}
\end{figure*}%

\begin{figure*}
\begin{center}
\includegraphics[width=179mm,clip]{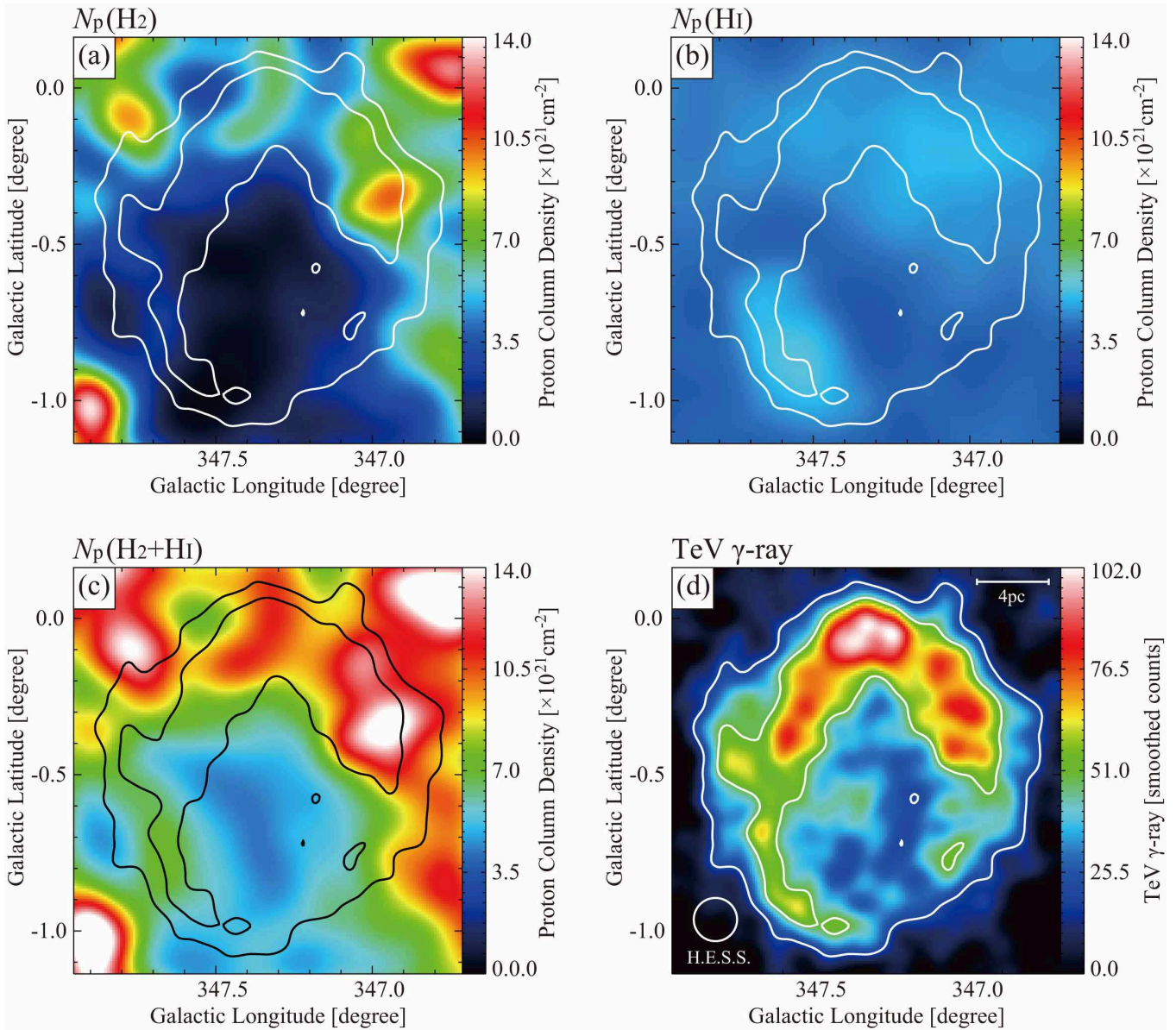}
\caption{(a) Distributions of column density of ISM protons $N_{\mathrm{p}}$ estimated from $^{12}$CO($J$=1--0) $N_{\mathrm{p}}$(H$_2$), (b) \ion{H}{1} emission with correction for the \ion{H}{1} self-absorption $N_{\mathrm{p}}$(\ion{H}{1}) and (c) sum of $N_{\mathrm{p}}$(H$_2$) and $N_{\mathrm{p}}$(\ion{H}{1}). All the datasets used here are smoothed to a HPBW of TeV $\gamma$-ray distribution with a Gaussian function. (d) TeV $\gamma$-ray distribution. Contours are plotted every 50 smoothed counts from 20 smoothed counts.}
\end{center}
\end{figure*}%

\begin{figure*}
\begin{center}
\figurenum{8}
\includegraphics[width=179mm,clip]{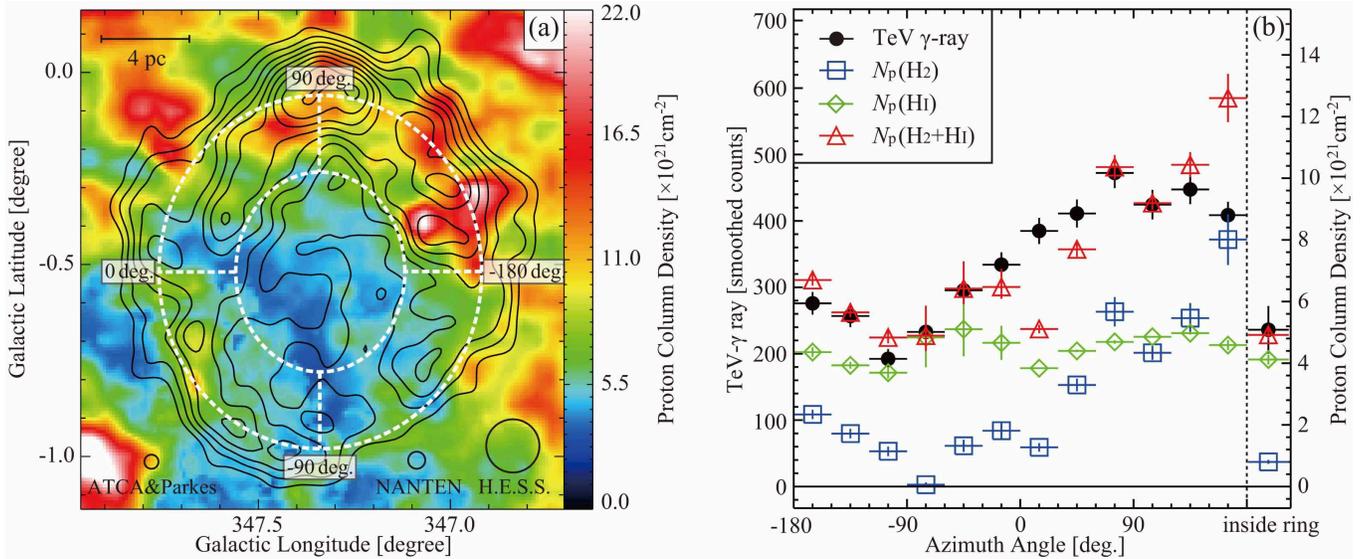}
\caption{(a) Distributions of column density of the total ISM protons $N_{\mathrm{p}}$(H$_2$+\ion{H}{1}) in a velocity range from $-$20 km s$^{-1}$ to 0 km s$^{-1}$. Contours are the same as in Figure 1(a). (b) Azimuthal distributions of $N_{\mathrm{p}}$(H$_2$), $N_{\mathrm{p}}$(\ion{H}{1}), $N_{\mathrm{p}}$(H$_2$+\ion{H}{1}) and TeV $\gamma$-ray smoothed counts per beam between the two elliptical rings shown in Figure 8(a). The proton column densities are averaged values between the rings (see text). Semi-major and semi-minor radii of the outer ring are 0.46 degrees and 0.42 degrees, respectively, and the radii of the inner ring are half of them. The same plots inside the inner ring are shown on the right side of Figure 8(b).}
\end{center}
\end{figure*}%

\begin{figure}
\begin{center}
\figurenum{9}
\includegraphics[width=86mm,clip]{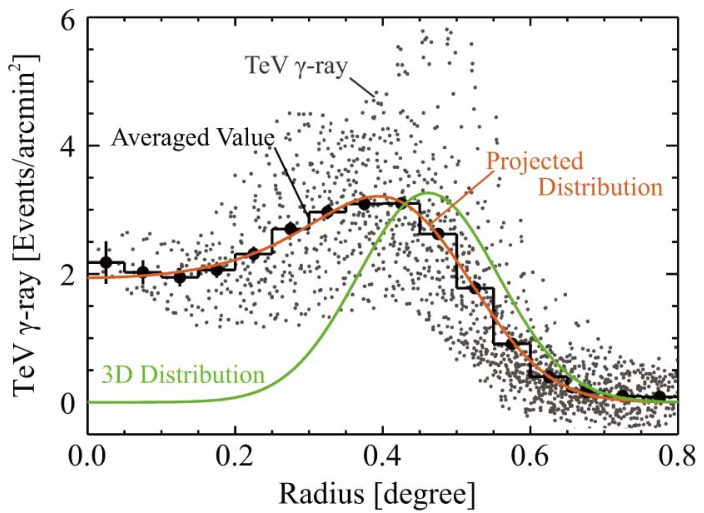}
\caption{Radial distribution of TeV $\gamma$-rays radiation. Small dots show the distributions of all the H.E.S.S. data points and large filled circles with error bars show averaged values at each radius. We assume a 3D spherical shell with a Gaussian-like intensity distribution along its radius to approximate the TeV $\gamma$-ray distribution (see text). The green line shows the estimated 3D Gaussian distribution and the red line shows its projected distribution. The peak radius and the full width at half maximum of the green line are estimated to be 0.46 deg ($\sim $ 8.0 pc) and 0.24 deg ($\sim $ 4.2 pc), respectively.}
\end{center}
\end{figure}%

We shall briefly review some basic properties of \ion{H}{1} gas in order to understand the behavior of \ion{H}{1} brightness \citep[e.g.][]{sato1978}. The spin temperature, $T_{\rm s}$, of \ion{H}{1} is $\sim $100 K or higher in warm neutral medium at particle density less than 10 cm$^{-3}$. $T_{\rm s}$ decreases with density from 100 K down to 10 K in a density range of 100--1000 cm$^{-3}$ (e.g., Figure 2 in \citeauthor{goldsmith2007} \citeyear{goldsmith2007}). The temperature decrease is mainly due to higher shielding of stellar radiation and increased line cooling. 

It is well established that \ion{H}{1} is converted into H$_2$ on dust surfaces with increasing of the gas column density and UV shielding and that H$_2$ is dissociated by cosmic rays and UV photons  \citep[e.g.,][]{allen1977}. The equilibrium \ion{H}{1} abundance is determined by the balance between formation and destruction of H$_2$ and the residual density of \ion{H}{1} is about 10$^{-2}$ that of H$_2$ in typical interstellar molecular clouds \citep{allen1977, sato1978}. We also note that the H$_2$ abundance should be time dependent since the formation of H$_2$ is a slow process in the order of 10 Myrs for density around 100 cm$^{-3}$ \citep[e.g.,][]{allen1977}.

Based on the \ion{H}{1}-H$_2$ transition, we interpret the dark \ion{H}{1} in Figure 2(b) as representing the \ion{H}{1} with lower $T_{\rm s}$. The CO W cloud shows a good spatial coincidence with the dark \ion{H}{1} W cloud as is consistent with the interpretation. The other prominent dark \ion{H}{1} region, the SE cloud, shows no CO and we suggest that its density is lower and its $T_{\rm s}$ is higher than in the CO W cloud. \ion{H}{1} brightness $T_L(V)$ is expressed as follows (e.g., Sato and Fukui 1978); 
\begin{eqnarray}
T_\mathrm{L}(V)=T_\mathrm{s}[1-e^{-\tau(V)}]+T^{\mathrm{FG}}_L(V)\nonumber\\
+[T^{\mathrm{BG}}_L(V)+T^{\mathrm{BG}}_C]e^{-\tau (V)} - (T^{\mathrm{FG}}_C+T^{\mathrm{BG}}_C)
\end{eqnarray}
where $T_\mathrm{L}(V)$, $T_\mathrm{s}$, $\tau (V)$, $T^{\mathrm{FG}}_L(V)$ and $ T^{\mathrm{BG}}_L(V)$ are the observed \ion{H}{1} brightness temperature, the spin temperature, the optical depth of cold \ion{H}{1} in the cloud, and the foreground and background \ion{H}{1} brightness temperature, respectively, at velocity $V$. $T^{\mathrm{FG}}_C$ and $T^{\mathrm{BG}}_C$ are the continuum brightness temperature at 21-cm wavelength  in the foreground and background of the cloud, respectively. The radio continuum emission is weak in RX J1713.7$-$3946 \citep{lazendic2004}, and  $T^{\mathrm{FG}}_C$ and $T^{\mathrm{BG}}_C$ are nearly zero as compared with $T_L(V)$.

We are then able to estimate the \ion{H}{1} column density of dark \ion{H}{1} clouds. Figure 4(b) shows the \ion{H}{1} self-absorption dip with the background \ion{H}{1} emission interpolated by a straight line connecting the two shoulders at 0 and 20 km s$^{-1}$. This gives a conservative estimate because the actual background \ion{H}{1} shape perhaps has a more intense peak at $-10$ km s$^{-1}$ as seen in the northern area of the SNR. The spin temperature $T_{\rm s}$ of the dark \ion{H}{1} gas is an unknown parameter. We estimate $T_{\rm s}$ to be less than $\sim$55 K from the lowest \ion{H}{1} brightness at the bottom of the dip in Figure 4(b) and higher than $\sim$20 K, where the temperature of the CO clouds is $\sim$10 K \citep{sano2010}. We estimate the absorbing dark \ion{H}{1} column density to be $N_{\mathrm{p}}$(\ion{H}{1}) = 1.0$\times $10$^{21}$ cm$^{-2}$ (optical depth = 0.8), 1.8$\times $10$^{21}$ cm$^{-2}$ (optical depth = 1.1) and 3.1$\times $10$^{21}$ cm$^{-2}${ (optical depth =1.5) for assumed three cases $T_{\rm s}$ = 30, 40 and 50 K, respectively, for the half-power line width $\Delta v $=10 km s$^{-1}$, where the \ion{H}{1} optical depth $\bar{\tau}$ is estimated by equation (2) and $N_{\mathrm{p}}$(\ion{H}{1}) by the following relationship;
\begin{eqnarray}
N_{\mathrm{p}}(\mathrm{H} \; \mathrm{I}) \; \mathrm{(cm^{-2})}=1.823\times10^{18}T_s \: \mathrm{(K)} \: \Delta v \mathrm{(km \; s^{-1})} \bar{\tau }
\end{eqnarray}
 We shall here adopt $T_{\rm s}$ = 40 K and a corresponding dark \ion{H}{1} optical depth of 1.1. A higher $T_{\rm s}$ gives a higher optical depth and vice versa. The relatively large optical depth around 1 is consistent with the fairly flat \ion{H}{1} dip in Figure 4(b), which suggests weak saturation. We also tested the effects of elevating the background \ion{H}{1} by 15 K and found a small change of 5 $\times $ 10$^{20}$ cm$^{-2}$. The error is mainly introduced by the straight-line approximation and uncertainty in $T_{\rm s}$ of $\sim$10 K. We infer the dark \ion{H}{1} column density is accurate within a systematic error of $\sim $1$\times $10$^{21}$ cm$^{-2}$.

The average \ion{H}{1} density in the SE cloud is roughly estimated to be 150 cm$^{-3}$ by dividing $1.8\times10^{21}$ cm$^{-2}$ by $\sim$4 pc, the line of sight length of the thick ISM shell, following the three-dimensional model described in Sub-section 3.5. This density is significantly lower than the critical density for collisional excitation of the CO ($J$=1--0) transition, $\sim$1000 cm$^{-3}$, consistent with no CO emission from the SE cloud and with low spin temperature around 40 K.

We also extended such an analysis to the regions with narrow \ion{H}{1} dips associated with CO emission, where we adopt $T_{\rm s}$ = 10 K, the kinetic temperature of the CO gas. The small dips in these regions indicate that the \ion{H}{1} optical depth is generally as low as $\sim$0.1 reflecting a small fraction of the residual \ion{H}{1} in CO gas. We show the distributions of the peak optical depth of the \ion{H}{1} self-absorption in {Figure 6(a)}, and the derived total \ion{H}{1} column density distribution, the sum of the \ion{H}{1} in emission and self-absorption in {Figure 6(b)}, where the SE cloud is significant. We shall hereafter refer to the dark \ion{H}{1} of $T_{\rm s}$ = 40 K as "cool \ion{H}{1}" and that of $T_{\rm s}$ = 10 K as "cold \ion{H}{1}".

\subsection{Total ISM protons}

The number of the total ISM protons in the SNR is given by summing up the three components in a velocity range from $-20$ to 0 km s$^{-1}$; H$_2$ derived from $^{12}$CO($J$=1--0), dark \ion{H}{1} (dips) and warm \ion{H}{1} (emissions). The results are shown as spatial distributions in {Figure 7}. Figure 7(a)--(d) show $N_{\mathrm{p}}$(H$_2$), $N_{\mathrm{p}}$(\ion{H}{1}), $N_{\mathrm{p}}$(H$_2$+\ion{H}{1}) and TeV $\gamma$-rays, respectively. We see the total ISM protons $N_{\mathrm{p}}$(H$_2$+\ion{H}{1}) shows a shell-like shape similar to the TeV $\gamma$-rays which significantly improves the correlation with the $\gamma $-rays as compared with the case of molecular gas only. We therefore conclude that the contribution of \ion{H}{1} is critical as well as H$_2$ in counting the ISM protons. We find that in the south the total ISM proton is dominated by the atomic gas, whereas in the north the molecular and atomic protons are both important. A more quantitative comparison will be given in Sub-section 3.5.2. Similar diagrams of the total ISM protons to Figure 7 are presented for the optically-thin case for reference Figure C1 in Appendix C, where the shell-like distribution toward the SE cloud is missing.

\subsection{The $\gamma$-rays and the ISM protons}

\subsubsection{Gamma-ray distribution}
The TeV $\gamma$-ray distribution obtained by H.E.S.S. is a nearly circular-symmetric shell with some ellipticity elongated in the north-south direction. In order to gain an insight into the distribution of the $\gamma$-ray emissivity we undertake a simple analysis of the $\gamma$-ray distribution. We first adopt an elliptical annular ring in the analysis, while \cite{aharonian2006b} made a similar analysis by using a circular annular ring in correlating $\gamma$-rays and NANTEN CO intensity (see their Figure 17). 

We estimated the radius of the $\gamma$-ray shell as defined at a half-intensity level of the peak $\gamma$-ray smoothed count every 15 degrees for an assumed center. We averaged the radii in angle and minimized the sum of the squares of the deviation from the average. This process gives a central position to be ($l$, $b$) = (347$\fdg$34, $-0\fdg$52). For this central position, we plotted the radius every 15 degrees and found that a sinusoidal distribution is a reasonable approximation as expected. Fitting this plot by a sinusoidal curve, we find the shell is approximated by an elliptical shape with an aspect ratio of 1.1 whose major axis is almost in the north-south direction. This elliptical shape is adopted in Figure 8(a).

Figure 9 shows the radial scatter of $\gamma$-ray smoothed counts and an averaged value shown by a step function in radius $r$ every 0.05 degrees. Here we also adopted the elliptical shape and normalized the radius to that of the major axis with the elliptical modification. After several trials of different functional forms, we found a Gaussian radial distribution of the $\gamma$-ray emissivity per volume reproduces well the projected radial distribution in Figure 9. In the fitting we have two free parameters of the Gaussian shape, the peak radius $r_0$ and the sigma $\sigma $ expressed as follows;
\begin{eqnarray}
F(r) = A \times e^{-\bigl(r - r_0\bigr)^2 / 2\sigma ^2}
\end{eqnarray}
where A is a normalization coefficient. By requiring that the error in the fitting becomes minimum in the projected distribution shown by the step function, we found $r_0$ = 0.46 degrees and $\sigma $ = 0.10 degrees give the best fit as shown in Figure 9. This distribution shows that the observed shell is consistent with a shell of a half-intensity thickness $\sim $0.24 degrees with nearly zero emissivity toward the center. This analysis indicates that the $\gamma$-rays are mainly emitted in a thick shell of 8.0 pc radius and 4.2 pc width at the half-intensity level with nearly zero emission from the inner part. A similar thick-shell model was also obtained by \cite{aharonian2006b}. Numerical modeling of the $\gamma$-ray emission has been undertaken by several authors and indicates that the $\gamma$-ray emission has a rather steep gradient beyond the peak of the shell in either of the leptonic or hadronic scenario \citep[e.g.][]{jun1996,zirakashvili2010}. The fitting to the H.E.S.S. data above shows that the gradient in the $\gamma$-ray distribution is not so steep toward the outside, which may be due to smearing in space by averaging. We shall not try a further elaborated analysis here due to the limiting angular resolution of H.E.S.S. which is 0.14 deg (FWHM).

Figure 9 shows that the projected radial distribution of ISM protons follows a fairly similar distribution to the $\gamma$-rays inside the SNR. This is consistent with that the ISM distribution is also shell-like with an inner cavity as is consistent with the stellar wind shell discussed in Sub-section 4.1; if the ISM has no cavity in the inner part, the projected distribution of the ISM should increase toward the center. We shall assume hereafter that the ISM distribution is also approximated by the same Gaussian shape as the $\gamma$-rays with a radius of 8.0 pc with a thickness of 4.2 pc at the half-intensity level.

\subsubsection{Comparison between the $\gamma $-rays and the ISM protons}

\begin{figure}
\begin{center}
\figurenum{10}
\includegraphics[width=86mm,clip]{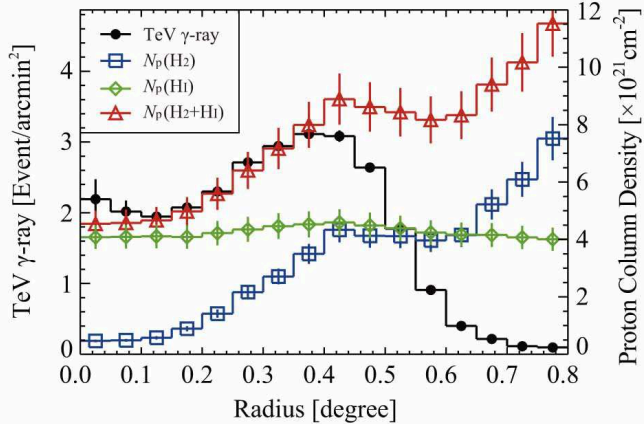}
\caption{Radial distributions of averaged values of TeV $\gamma$-rays radiation, $N_{\mathrm{p}}$(H$_2$), $N_{\mathrm{p}}$(\ion{H}{1}) and $N_{\mathrm{p}}$(H$_2$+\ion{H}{1}). $N_{\mathrm{p}}$(H$_2$) and $N_{\mathrm{p}}$(\ion{H}{1}) show column densities estimated from $^{12}$CO($J$=1--0) and \ion{H}{1}, respectively, and $N_{\mathrm{p}}$(H$_2$+\ion{H}{1}) shows the total ISM proton column density.}
\end{center}
\end{figure}%

In the hadronic scenario, the target distribution should be correlated with the $\gamma$-ray distribution for a uniform CR distribution. This correlation should be seen inside the shell of the SN shock which has a sharp gradient beyond its outer radius. We expect that the ISM protons are distributed beyond the outermost edge of the shell where CR protons cannot reach by diffusion. Beyond the SNR shock, the $\gamma$-ray emission profile may be influenced by components from the diffuse cosmic-ray background and by the energy dependent transport of escaping cosmic-rays from RX J1713.7$-$3946 into the clumpy ISM \citep[e.g.,][]{gabici2009, casanova2010a,casanova2010b}. We are able to avoid possible effects of such cut-off by taking the radius of the correlation analysis well within the SNR shell where the CR protons do not decrease in energy density. 

The ISM proton distribution is shown in Figure 8(a) with the two annular elliptical rings along the shell, where the size of the outer ring was chosen to meet the requirement above. Figure 8(b) shows a comparison between the ISM protons and $\gamma$-rays in the position angle shown in Figure 8(a), where the vertical scale is adjusted so that the correspondence with the TeV $\gamma$-rays becomes optimum. Here, the error in the TeV $\gamma$-ray emission from the publicly available H.E.S.S. image is approximately (smoothed counts)$^{0.5}$. In Figure 8(b) the uncertainty in the dark \ion{H}{1} in the SE cloud, 1$\times $10$^{21}$ cm$^{-2}$, is in the order of 10--20 $\%$ of the total. The total ISM proton density shows a good agreement with the TeV $\gamma$-ray angular distribution and also the central part in the inner ring. We recall that CO alone showed marked deficiency toward the SE cloud as compared with the $\gamma$-rays (see Figure 17 of \citeauthor{aharonian2006b} \citeyear{aharonian2006b}). The present analysis indicates the deficiency is recovered by including \ion{H}{1} and has shown that the total gas of both atomic and molecular components have a good correlation with the TeV $\gamma$-rays in the annular ring. The total mass of the ISM protons responsible for the $\gamma$-rays is 2.0$\times 10^4$ $M_{\odot}$ over the whole SNR (radius 0.65 deg); the mass of molecular protons is 0.9$\times 10^4$ $M_{\odot}$ and that of atomic protons is 1.1$\times 10^4$ $M_{\odot}$, where we assume the ISM protons interacting with the CR protons is proportional to the TeV $\gamma$-rays (Sub-section 3.5.1., Figure 10).

There are two points in Figure 8(b), for which additional remarks may be appropriate. One is the point at an azimuth angle of 115 degrees which may be estimated too low due to lack of correction for the self-absorption because of the large velocity shift in the expanding shell (see Figure B1). Another is the point at an azimuth angle of 165 degrees where the strong CO emission (peak A after \citeauthor{fukui2003} \citeyear{fukui2003}) increases the proton column density, although the increased protons may not be interacting with the CR protons beyond the SNR shock, leading to less $\gamma $-rays. 

An independent test is made by the radial distribution of the ISM protons given in Figure 10, where an average taken over the same binning as the $\gamma$-rays in Figure 9 is shown by a step function {and} the total ISM protons and $\gamma$-rays are superposed with the same proportional factor as adopted in Figure 8. Here, the error in the TeV $\gamma$-ray emission is approximately (oversampling-corrected total smoothed count)$^{0.5}$ normalized to 1 (arcmin)$^2$. We see the $N_{\mathrm{p}}$(H$_2$+\ion{H}{1}) and $\gamma$-rays show a good agreement inside the shell and the $\gamma$-rays sharply decrease outside the shell. This offers another presentation of the good correlation between the $\gamma$-rays and the ISM protons.

We argue that the apparent anti-correlation between the \ion{H}{1} brightness at the bottom of the dips and the $\gamma$-rays in the SE cloud (Figure 4) is consistent with that the \ion{H}{1} dips are due to the cool and dense \ion{H}{1} gas. The anti-correlation is interpreted that the spin temperature $T_{\rm s}$ of \ion{H}{1} decreases with density (Sub-section 3.3) and that the $\gamma$-rays increase with the ISM proton density locally in the SE cloud, demonstrating detailed correspondence between the $\gamma$-rays and the ISM protons which is mainly atomic. The small and narrow \ion{H}{1} dips in the W and N clouds have the \ion{H}{1} column density less than 10$^{20}$ cm$^{-2}$, significantly lower than the typical molecular column density by two orders of magnitude. So, in most of the regions except for the SE cloud the \ion{H}{1} column density is dominated by emission but not by self-absorption. For the sake of reference, we show a set of similar diagrams of ISM proton distributions for the optically-thin case in Figures C2 and C3 in Appendix C, corresponding to Figures 8 and 10, respectively.

\begin{deluxetable*}{lcc}
\tablecaption{A Comparison between RX J1713.7$-$3946 and Pegasus Loop}
\tablewidth{0pt}
\label{table3}
\tablehead{& RX J1713.7$-$3946$^{\ast}$ & Pegasus Loop$^{\dagger}$}
\startdata
Distance (kpc)......................................................................& 1 & 0.1 \\
Diameter (pc).......................................................................& 17.4 & 25 \\
Total mass of the ISM ($M_{\mathrm{\odot}}$)................................................ & $\sim$20000$^{\ddagger}$ & $\sim $1500 \\
Thickness of the ISM shell (pc)............................................ & $\sim$4.2 & $\sim$5 \\
Peak brightness of \ion{H}{1} (K).................................................... & $\sim$170 & $\sim$40 \\
Linewidth of \ion{H}{1} (km s$^{-1}$).................................................... & $\sim$20 & $\sim$16 \\
Expansion velocity the gaseous shell (km s$^{-1}$).................... & $\sim$10 & $\sim$7--9 \\
Spectral type of the projenitor.............................................. & B1 V / B0 V$^{\ast \ast}$ & B2 IV \\
\enddata
\label{tab1}
\tablecomments{$^{\ast}$ \cite{fukui2003}, \cite{moriguchi2005}, $^{\dagger}$ \cite{yamamoto2006}, $^{\ast \ast}$ \cite{cassamchenai2004}, $^{\ddagger}$ present work. The Pegasus loop may consist of two shells and the the mass should be regarded as an upper limit \citep{yamamoto2006}.}
\end{deluxetable*}

Before concluding this Sub-section, we cautiously note that the cool/cold \ion{H}{1} could not be estimated accurately, if the cool/cold \ion{H}{1} is optically thick, if the cool/cold \ion{H}{1} lies behind optically thick foreground \ion{H}{1} in the line of sight, or if the background \ion{H}{1} profile has a different shape from its neighbors. Such effects, while posing intrinsic limits for probing cool/cold \ion{H}{1}, are relatively unimportant for nearby objects at a distance of 1 kpc or less where foreground \ion{H}{1} is not important. The dark \ion{H}{1} W and SE clouds are probably good examples where the cool/cold \ion{H}{1} is well traced by the low \ion{H}{1} brightness, whereas the N cloud with higher \ion{H}{1} brightness may be partially affected by the foreground \ion{H}{1} in the line of sight. 

\section{Discussion}\label{section:discussion}

\subsection{The evacuated cavity by the stellar wind}
It is likely that the CO shell in Figure 1(b) was formed over a timescale of Myrs by the stellar wind of the projenitor, an OB star that exploded as a supernova (SN) 1600 yr ago. The total velocity span of the CO shell, $\sim $20 km s$^{-1}$, is much smaller than the SN shock speed and indicates that it takes Myr to form the shell of the ISM as roughly estimated by dividing the radius 9 pc by 10 km s$^{-1}$. Molecular gas expanding at 10 km s$^{-1}$ can move only 0.01 pc in 1000 yrs. Therefore, the current CO distribution has little been affected by the supernova explosion (SNe) and holds the initial condition before the shock interaction. 

While a stellar-wind shell with a known central star is not often observed elsewhere, one such example is the Pegasus loop found in $^{12}$CO($J$=1--0), \ion{H}{1} and dust emission at ($l$, $b$) = (109$^{\circ }$, $-$45$^{\circ }$) centered on a run-way star HD886 (B2 IV) \citep{yamamoto2006}. The Pegasus loop is located at $\sim $100 pc in a relatively uncontaminated environment outside the Galactic plane. No SNe occurred yet in this shell. A comparison between RX J1713.7$-$3946 and the Pegasus loop is given in Table 1. In Pegasus the swept-up shell of the ISM has a width of $\sim $5 pc for a radius of $\sim $18 pc and a total mass of $\sim $1500 M$_{\odot }$. The shell is mostly atomic and consists of 78 smaller $^{12}$CO($J$=1--0) clumps (see Figure 10 in \citeauthor{yamamoto2006} \citeyear{yamamoto2006}). The clumped CO is a natural outcome of thermal/gravitational instability and seems common in such a shell. The shell is expanding at a total velocity span of 15 km s$^{-1}$. The \ion{H}{1} density inside the shell is $\sim $1 cm$^{-3}$ in the north, where the stellar wind evacuated the ISM over 1 Myrs. The Pegasus loop is located in a somewhat lower-density environment than RX J1713.7$-$3946 and offers an insight into the initial condition of the ISM prior to the SNe in RX J1713.7$-$3946.

Inoue, Yamazaki $\&$ Inutsuka (\citeyear{inoue2009}) and IYIF2011 carried out numerical simulations of the hydro-dynamical interaction between the shock wave and the highly inhomogeneous neutral gas to model the interaction in RX J1713.7$-$3946. The SN in RX J1713.7$-$3946 exploded in the cavity with average density less than 1 cm$^{-3}$ \citep[e.g.][]{zirakashvili2010,morlino2009,berezhko2008} and the dense shell with CO clumps remaining more or less as they were prior to the SNe. The SN shock front moves almost freely at $\geq 3000$ km s$^{-1}$ in the cavity in the early phase of $\sim$1000 yrs and begins to interact with the dense and thick clumpy ISM wall swept-up by the stellar wind only in the last few 100 yrs. The $\gamma$-ray shell is not strongly deformed, while we see some deviations of a pc scale from a perfect circular shell, suggesting effects of recent dynamical interaction.

The interaction between molecular clumps and the shock is observed as the X-ray enhancement around dense molecular clumps at a spatial resolution higher than 0.5 pc. 
\cite{sano2010} showed that the molecular clump peak C is rim-brightened in X-rays, suggesting that it is a dense clump overtaken by the shock, and peak A \citep{fukui2003} is also X-ray brightened only toward its inner edge, indicating the shock interaction at the inner boundary of peak A. IYIF2011 showed that the initial magnetic field $B$ of 1 $\mu$G is amplified to 0.1 to 1 mG near dense clumps by the enhanced turbulence driven by the shock. The stronger magnetic field explains the X-ray enhancement as due to the enhanced synchrotron emission that is proportional to $B^2$, or, due to increased acceleration. IYIF2011 also showed that the shock speed $v_s$ is significantly reduced locally with density $n$ (cm$^{-3}$) such that $v_s \sim $3000 km s$^{-1}$/$\sqrt{n / n_0}$, where $n_0$ =1 cm$^{-3}$. This dependence of $v_s$ on density can explain the absence of thermal X-rays in the SNR because the molecular gas is too dense to be affected by the shock to emit thermal X-rays (IYIF2011). A uniform lower-density case with significant thermal X-rays by shock heating is presented by \cite{ellison2010} but such a model is not applicable to the highly inhomogeneous ISM of RX J1713.7$-$3946 (IYIF2011, see also discussion in Section 4 of \citeauthor{ellison2010} \citeyear{ellison2010}). The picture above is also consistent with that peak C, having density greater than 10$^4$ cm$^{-3}$, has survived without erosion \citep{sano2010}. 

\subsection{The $\gamma$-ray emission mechanism}
TeV $\gamma$-rays are emitted via two mechanisms, either leptonic or hadronic processes. The leptonic process explains $\gamma$-rays via the inverse Compton effect between CR electrons and low energy photons. In the hadronic scenario $\gamma$-rays are emitted by the decay of neutral pions which are produced in the high energy reactions between CR protons and ISM protons. Diffusive shock acceleration (DSA) is the most widely accepted scheme of particle acceleration \citep{bell1978,blandford1978,jones1991,malkov2001}. The previous works on RX J1713.7$-$3946 show that the observed spectral energy distribution of $\gamma$-rays and X-rays is explained by either of the leptonic and/or hadronic mechanisms if DSA works to accelerate the particles \citep{aharonian2006b, porter2006, katz2008, berezhko2008, ellison2008, tanaka2008, morlino2009, acero2009, ellison2010, patnaude2010, zirakashvili2010, abdo2011, fang2011}

In the hadronic scenario, where the neutral pion decay determines the $\gamma$-rays via proton-proton reactions, the average density of the target protons is constrained by the total energy of CR protons; the average target density greater than 0.1 cm$^{-3}$ is required to produce CR protons having the total energy of $10^{51}$ erg, for the maximum energy of a SNe, while higher target density is required for less CR proton energy. In the leptonic scenario, where the inverse Compton process produces $\gamma$-rays, the critical parameter is the magnetic field which constrains the synchrotron loss timescale of CR electrons; a magnetic field of order of 10 $\mu $G is usually required \citep[e.g.][]{tanaka2008}.

We here argue that the highly inhomogeneous distribution of the ISM, the cavity and the dense and clumpy wall opens a possibility to accommodate the low-density site for DSA and the high-density target simultaneously as discussed into detail by IYIF2011. A similar argument on the hadronic interaction between CR protons with the ambient dense clouds has been presented by \cite{zirakashvili2010}. In this picture, first, the cosmic rays are accelerated via DSA in the low density cavity, and second, the CR protons reach and react with the target protons in the dense wall to produce $\gamma$-rays. The main energy range of the CR protons required for hadronic TeV $\gamma$-rays is 10--800 TeV \citep{zirakashvili2010}. The penetration depth, $l_{\rm pd}$, of cosmic rays is expressed as follows (IYIF2011);
\begin{eqnarray}
\nonumber l_{\rm pd} \sim 0.1\eta^{1/2} \left(\frac{E}{10\ {\rm TeV}}\right)^{1/2} \left(\frac{B}{100 \ {\rm \mu G}}\right)^{-1/2} \left(\frac{t_{\rm age}}{10^3 \ {\rm yr}}\right)^{1/2} \\ {\rm (pc)} \;\;\;\;\;\;
\end{eqnarray}
where $E$, $B$ and $t_{\rm age}$ are the particle energy, the magnetic field and the age of the SNR. The parameter $\eta$ is the so-called "gyro-factor" and has some ambiguity. In the SNR, it is reasonable to consider $\eta$ $\sim$1 at least around the cloud 
\cite{uchiyama2007}. Thus, the penetration depth of the protons in the above energy range is 0.3--2.8 pc for magnetic field of 10 $\mu$G and 0.1--0.9 pc for 100 $\mu$G in a typical timescale of $\sim$$10^3$ yr. The penetration depth of the CR electrons is determined by taking $t_{\mathrm{age}}$ equal to the synchrotron loss timescale \citep[e.g.][]{tanaka2008} in equation (6) and becomes energy-independent for the X-ray emitting electrons of 1--40 TeV as follows;
\begin{eqnarray}
l=0.026 \ \eta^{1/2} \left(\frac{B}{100 \ {\rm \mu G}}\right)^{-3/2}\ \ {\rm (pc)}
\end{eqnarray}
We estimate $l$ to be from 0.8 pc for 10 $\mu$G to 0.026 pc for 100 $\mu$G if $\eta=1$. CR protons can therefore reach and penetrate into the dense gas within pc-scale of the acceleration site to produce TeV $\gamma$-rays, while the CR electrons stay relatively closer to the acceleration site, in particular, near the dense gas having strong magnetic field. This offers an explanation on the hadronic $\gamma$-ray production and the correlation between the $\gamma$-rays and target protons in Figures 4, 8 and 10 is a natural outcome in the scenario (IYIF2011).

\cite{gabici2007} discussed the importance of the energy-dependent interaction between CR protons and molecular clouds and \cite{zirakashvili2010} discussed that the $\gamma$-ray spectrum may not distinguish the leptonic and hadronic scenarios in case of RX J1713.7$-$3946 due to such energy dependence. Recently, Fermi-LAT observations showed that the GeV spectrum of RX J1713.7$-$3946 is hard, similar to what is expected in the leptonic scenario, and  \cite{abdo2011} discussed that the hard spectrum may favor to the leptonic scenario. IYIF2011, however, argued that the hard Fermi-LAT GeV spectrum is explained well also by the hadronic scenario as due to the energy-dependent penetration of CR protons into the dense clouds and that the leptonic scenario is not unique to explain the spectrum. IYIF2011 confirmed that the $\gamma$-ray spectrum becomes similar both for the leptonic and hadronic scenarios, not usable to distinguish the two scenarios, as noted by \cite{zirakashvili2010} and concluded that the hadronic origin is testable only by comparing $\gamma$-rays with the ISM target distribution. The present results have demonstrated that the ISM proton distribution show indeed a good spatial correspondence with the $\gamma$-rays by taking into account the contribution of the \ion{H}{1} and match with the prediction by \cite{zirakashvili2010} and IYIF2011. 

The total energy of CR protons is estimated by the relationship between the total target protons and the observed $\gamma$-rays (2--400 TeV) after extrapolating the proton spectrum to 1 GeV as follows \citep{aharonian2006b}; 
\begin{eqnarray}
{W_\mathrm{tot} \sim 1-3 \times 10^{50} \biggl(\frac{d}{1\, \mathrm{kpc}} \biggr)^2 \biggl(\frac{n}{1\, \mathrm{cm^{-3}}}\biggr )^{-1}} \ \ {\rm (erg)}
\label{equation:wtot}
\end{eqnarray}
where the distance to the source $d \sim$1 kpc and the density of the target protons is $n$. The average density of ISM protons is calculated to be $\sim $130 cm$^{-3}$ for the total mass of the ISM protons 2.0$\times 10^4$ $M_{\odot}$ over the whole SNR (radius 0.65 degrees) as modeled in Figure 10 and the total CR proton energy to be $\sim$0.8--2.3 $\times$ 10$^{48}$ ergs by using equation (\ref{equation:wtot}). This corresponds to $\sim$ 0.1 \% of the total energy release of a SNe and may appear low. The other SNRs like W44 and W28 of a few to 10 times 1000 yrs old have the total CR proton energy in the order to 10$^{49}$--10$^{50}$ ergs \citep{abdo2010,giuliani2010}. We may speculate that the CR protons become accumulated in a few times 10000 yrs to reach more than 10 \% of the SNe energy. This issue is to be further tested by examining cosmic ray escaping from SNRs \citep[e.g.][]{gabici2009,casanova2010a,casanova2010b}.

To summarize the discussion, we have shown that a combined analysis of CO and \ion{H}{1} provides a reasonable candidate for the target ISM protons and thereby lends a new support for the hadronic scenario. We should note that the present analysis offers one of the necessary conditions for the hadronic scenario for uniform CR proton distribution, but it is not a full verification of the hadronic scenario and does not rule out leptonic components. We need to acquire additional observations before fully establishing the hadronic scenario, including better determination of the magnetic field and higher angular resolution images of $\gamma$-rays at least comparable to that of the ISM. Cherenkov Telescope Array will provide such images in future. We discussed that the observed highly inhomogeneous distribution of the ISM plays an essential role in the $\gamma$-ray production; DSA works in highly evacuated cavity and the accelerated CR protons travel over a pc to interact with the surrounding dense ISM protons. It is important to develop a similar analysis of both \ion{H}{1} and CO in the other similar objects like RX J0852.0-4622 (Vela Jr.), RCW86 and HESS J1731-347. Such works are in progress based on the NANTEN2 observations and high-resolution \ion{H}{1} interferometry.

\section{Conclusions} \label{section:conclusions}
We summarize the main conclusions as follows;

\begin{enumerate}
\item A new analysis of CO and \ion{H}{1} has revealed that the TeV $\gamma$-ray SNR RX J1713.7$-$3946 is associated with a significant amount of \ion{H}{1} gas without H$_2$ derived from CO. This \ion{H}{1} gas is relatively dense and cold and detectable mainly as \ion{H}{1} emission. We have also identified regions where \ion{H}{1} is observed as dark \ion{H}{1} in self-absorption dips and derived the total ISM proton column density over the SNR. The \ion{H}{1} plus H$_2$, the total ISM protons, provides one of the necessary conditions, target protons,  in the hadronic origin of the $\gamma$-rays. Such target ISM protons have not been identified in the previous study that took into account only H$_2$, although the present finding alone does not exclude the leptonic origin.

\item For an annular pattern around the TeV $\gamma$-ray shell, we compared the total ISM proton distribution with the TeV $\gamma$-ray distribution and found that they show reasonably good correspondence, varying by similar factors. The inclusion of the atomic protons observed as the \ion{H}{1} self-absorption dips is essential particularly in the southeast of the $\gamma$-ray shell. The interpretation of \ion{H}{1} self-absorption dips is also supported by the enhanced optical extinction toward the southeast rim.

\item The cavity surrounding the SNR was created by the stellar wind of the SN projenitor. The inside of the cavity is of low density with $<$1 cm$^{-3}$ while the cavity wall consists of the dense and clumpy atomic or molecular target protons of $\geq $100--1000 cm$^{-3}$. The diffusive shock acceleration in the highly inhomogeneous ISM offers a reasonable mechanism of particle acceleration in the low-density cavity and the dense wall acts as the target for $\gamma$-ray production by the CR protons. Hydro-dynamical numerical simulations of the interaction have shown detailed physical processes involved (IYIF2011).

\item By considering the other pieces of the observational and theoretical works accumulated thus far, the present results make the hadronic interpretation much more comfortable in RX J1713.7$-$3946. The current energy of the total CR protons is estimated to be $\sim $10$^{48}$ ergs, 0.1 $\%$ of the total energy of SNe, if we assume the $\gamma$-rays are all produced by the hadronic process. 
\end{enumerate}

\acknowledgments
We are grateful to Felix Aharonian for the lively and fruitful discussion on the subject, without which the work would not have been completed. The NANTEN project is based on a mutual agreement between Nagoya University and the Carnegie Institution of Washington (CIW). We greatly appreciate the hospitality of all the staff members of the Las Campanas Observatory of CIW. We are thankful to many Japanese public donors and companies who contributed to the realization of the project. NANTEN2 is an international collaboration of ten universities, Nagoya University, Osaka Prefecture University, University of Cologne, University of Bonn, Seoul National University, University of Chile, University of New South Wales, Macquarie University, University of Sydney and ETH Zurich. The work is financially supported by a grant-in-aid for Scientific Research (KAKENHI, no. 15071203, no. 21253003, no. 20244014, no. 23403001, no. 22540250, and no. 22244014) from MEXT (the Ministry of Education, Culture, Sports, Science and Technology of Japan) and JSPS (Japan Society for the Promotion of Science) as well as JSPS core-to-core program (no. 17004). We also acknowledge the support of the Mitsubishi Foundation and the Sumitomo Foundation. This research was supported by the grant-in-aid for Nagoya University Global COE Program, $"$Quest for Fundamental Principles in the Universe: from Particles to the Solar System and the Cosmos$"$, from MEXT. The satellite internet connection for NANTEN2 was provided by the Australian Research Council. 

\appendix
\section*{APPENDIX A\\ Velocity channel distributions in RX J1713.7$-$3946}
We show velocity channel distributions of $^{12}$CO($J$=1--0, 2--1) and \ion{H}{1} every 1 km s$^{-1}$ from $-$20 km s$^{-1}$ to 0 km s$^{-1}$ superposed on the TeV $\gamma$-ray distribution in Figure A. 

\clearpage
\begin{figure*}
\begin{center}
\figurenum{A}
\includegraphics[width=179mm,clip]{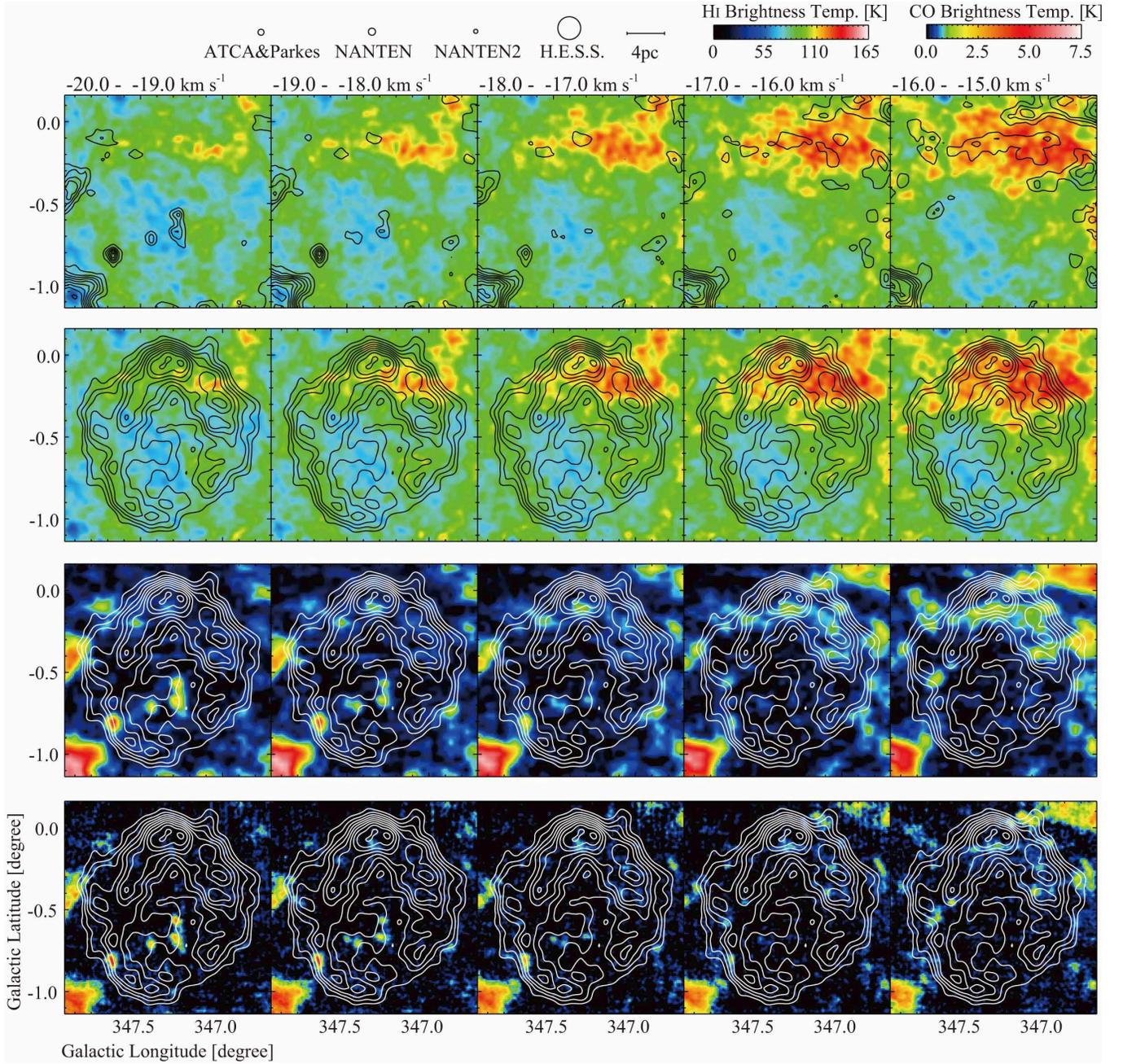}
\caption{Velocity channel distributions of $^{12}$CO($J$=1--0, 2--1) emission and \ion{H}{1} brightness temperature overlayed on the TeV $\gamma$-ray distribution. $First$-$row$ $panels$ $(top)$: \ion{H}{1} image and $^{12}$CO($J$=1--0) contours. $Second$-$row$ $panels$: \ion{H}{1} image superposed on the TeV $\gamma$-ray contours. $Third$-$row$ $panels$: $^{12}$CO($J$=1--0) image superposed on the TeV $\gamma$-ray contours. $Fourth$-$row$ $panels$: $^{12}$CO($J$=2--1) image superposed on the TeV $\gamma$-ray contours. Each panel shows CO and \ion{H}{1} distributions every 1 km s$^{-1}$ in a velocity range from $-$20 to 0 km s$^{-1}$. The lowest contour levels of CO and TeV $\gamma$-rays are 0.73 K ($\sim 3\sigma $) and 20 smoothed counts and contour intervals of CO and TeV $\gamma$-rays are 0.73 K ($\sim 3\sigma $) and 10 smoothed counts, respectively.}
\label{a1a2}
\end{center}
\end{figure*}%

\begin{figure*}
\begin{center}
\figurenum{A}
\includegraphics[width=179mm,clip]{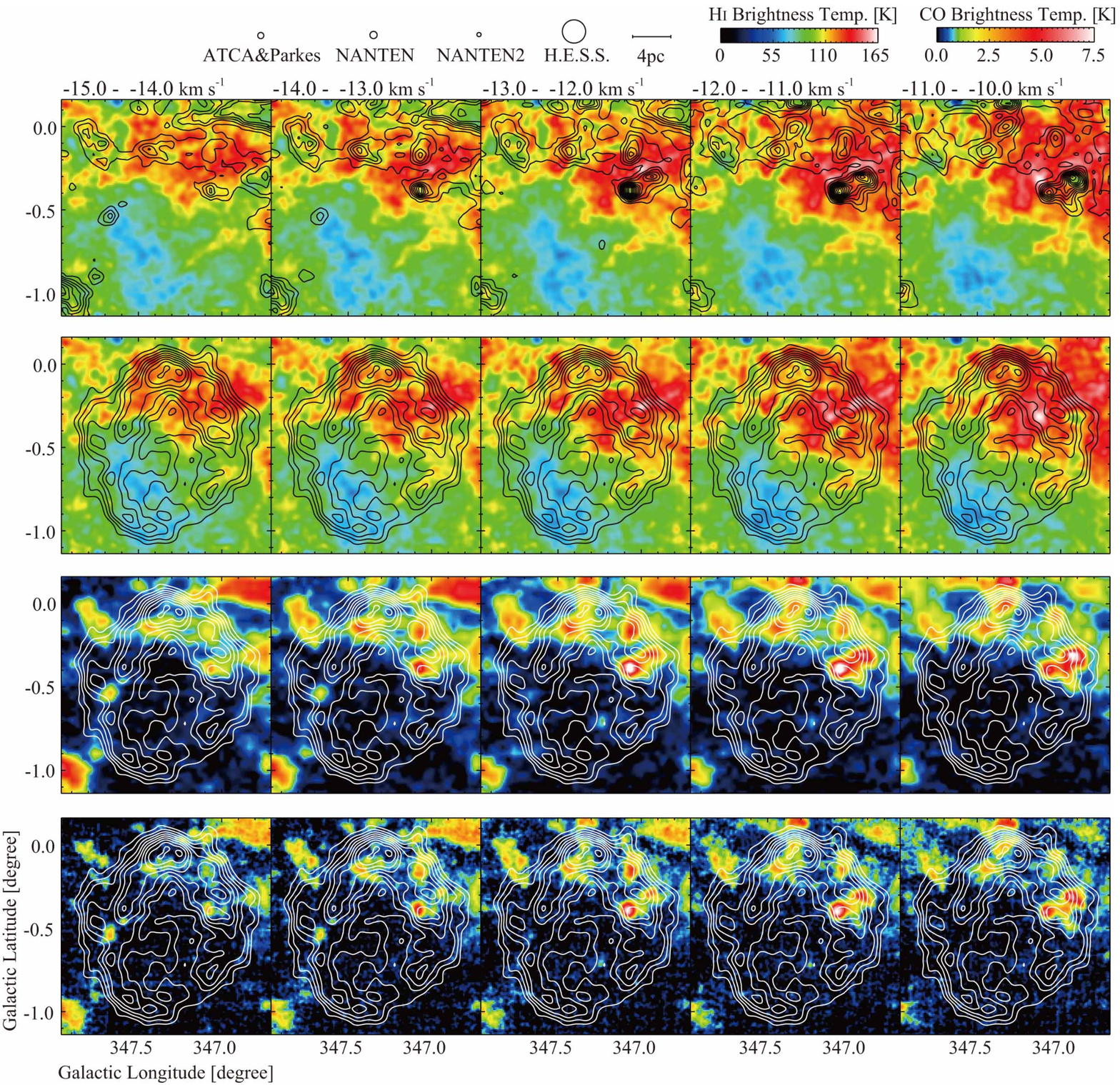}
\caption{$Continued$}
\label{a1b2}
\end{center}
\end{figure*}%

\begin{figure*}
\begin{center}
\figurenum{A}
\includegraphics[width=179mm,clip]{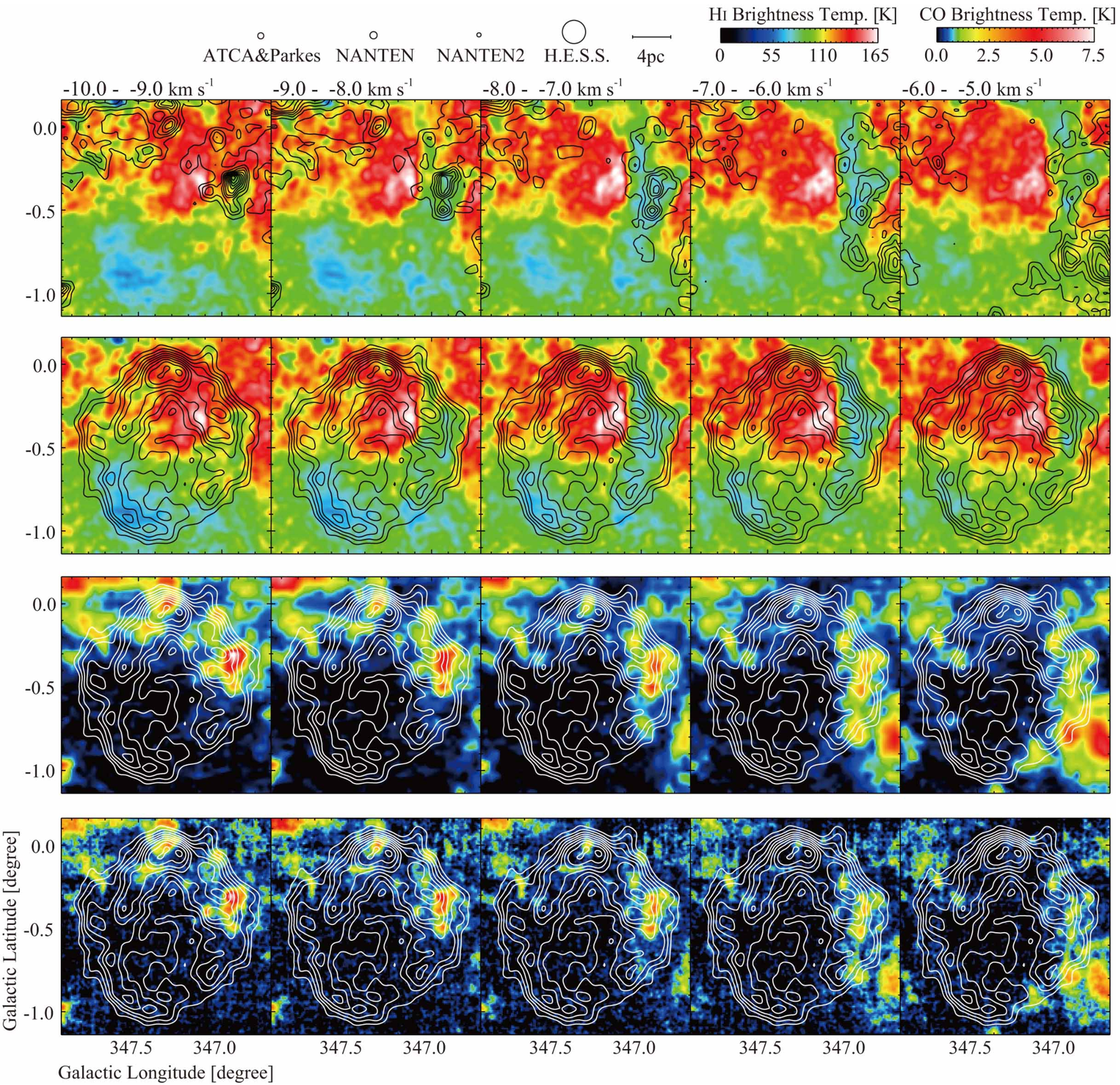}
\caption{$Continued$}
\label{a1c2}
\end{center}
\end{figure*}%

\begin{figure*}
\begin{center}
\figurenum{A}
\includegraphics[width=179mm,clip]{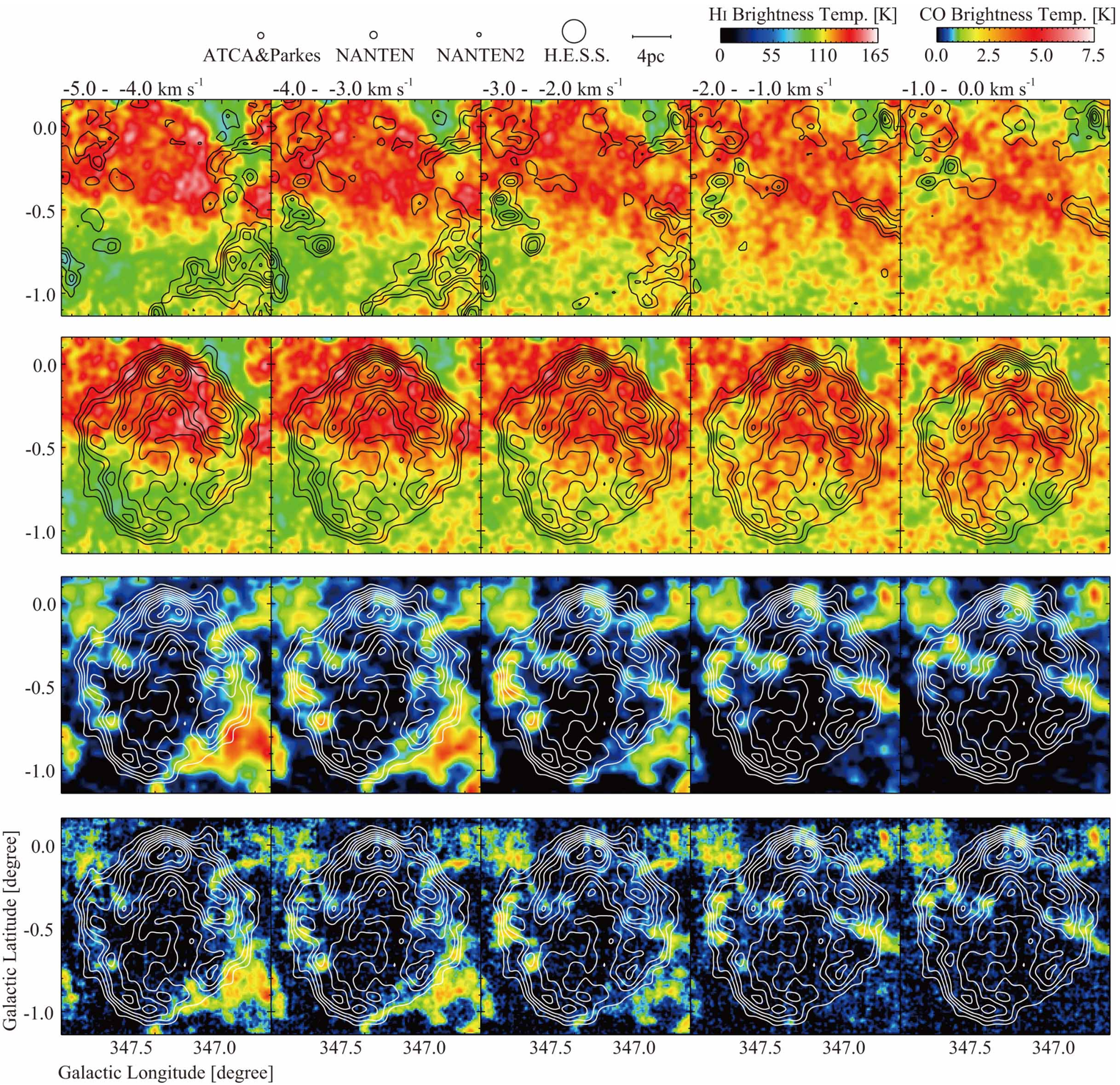}
\caption{$Continued$}
\end{center}
\end{figure*}%

\section*{APPENDIX B\\ Expanding motion of the dark \ion{H}{1} SE cloud}
Figure B1 left shows schematically an expanding spherical shell of radius $R_0$ = 9 pc and uniform expansion velocity $V_0$ = 10 km s$^{-1}$ and Figure B1 right a position-velocity diagram of the shell, where the ellipsoidal nature of the shell is not taken into account for simplicity. 
Figure B2 shows three representative velocity-channel distributions of the dark \ion{H}{1} SE cloud for a velocity range from $-20$ to $-10$ km s$^{-1}$ and shows that the SE cloud is extended to the north. The extension shifts toward the northwest with velocity decrease from $-10$ to $-20$ km s$^{-1}$ as is consistent with the iso-velocity contours expected from the shell model in Figure B1. 
Figure B3 shows another presentation of kinematical details of the SE cloud in position-velocity diagrams. We choose a line AB passing through the center of the SNR and the SE cloud, and another line CD passing through the SE cloud in the north-south (Figure B3(a)). We show a position-velocity distribution of \ion{H}{1} along the line AB (Figure B3(b)) and \ion{H}{1} profiles along the two lines AB and CD (Figure B3(c)). We find the SE cloud is extended to the northwest with a large velocity gradient of 10 km s$^{-1}$ per 0.5 degrees, or $\sim $1.2 km s$^{-1}$ pc$^{-1}$. The \ion{H}{1} profiles in Figure B3(c) shows that the dips are deep and clear at $b$ less than $-0.5$ degrees but becomes shallower above $b$ = $-0.5$ degrees. The shallower dips make it nontrivial to quantify the dips at $b$ higher than $-0.5$ degrees; we note that, even when the dips are not clearly seen, the \ion{H}{1} probably suffers from self-absorption to some extent as suggested by the weaker \ion{H}{1} brightness at $-12$ km s$^{-1}$ toward $b$ = $-0\fdg52$ than toward $b$ = $-0\fdg35$ (line AB).
We note that the strong velocity gradient in Figures B2 and B3 is consistent with the blue-shifted part of an expanding shell. The strong velocity gradient is interpreted in terms of the expanding shell as depicted by a white circle in the position-velocity diagram (Figure B3(b)). The blue shift by 10 km s$^{-1}$ toward the center of the SNR indicates this part of the shell is in the foreground. This is consistent with that the dips are due to self-absorption against the background \ion{H}{1} emission. We also infer that the swept-up shell is highly non-uniform since the broad \ion{H}{1} dips are seen only in a quarter of the shell.

\section*{APPENDIX C\\Analysis of the \ion{H}{1} emission; the optically thin case}
The present analysis has shown that the \ion{H}{1} is self-absorbed in part of the SNR as indicated by the \ion{H}{1} dips and the \ion{H}{1} column density is estimated by taking into account the self-absorption (Figure 7). In order to see the effects of the self-absorption quantitatively, we here show for comparison the ISM proton distribution in the optically thin case, which does not take into account the self-absorption.
Figure C1, equivalent to the self-absorption case in Figure 7, includes the \ion{H}{1} column density distribution for the optically-thin assumption smoothed to the HESS resolution ((b) and (c)), where the SE cloud is not seen. Figures C1(a) and (d) are the same with those in Figure 7.
Figure C2 is equivalent to Figure 8. Figure C2(a) is the total ISM proton column density for the optically thin \ion{H}{1} at NANTEN resolution overlayed on the TeV $\gamma$-ray distribution. Figure C2(b) is the corresponding azimuthal distribution of ISM protons and TeV $\gamma$-rays, where the ISM protons is deficient in azimuthal angle from $-90$ to 0 degrees as compared to Figure 8(b).
Figure C3 is equivalent to Figure 10, and shows the radial distribution of ISM protons for the optically thin \ion{H}{1} without correction for the \ion{H}{1} self-absorption. In the smoothed radial distribution, the effect of the self-absorption is not so obvious. 

\clearpage
\begin{figure*}
\begin{center}
\figurenum{B1}
\includegraphics[width=160mm,clip]{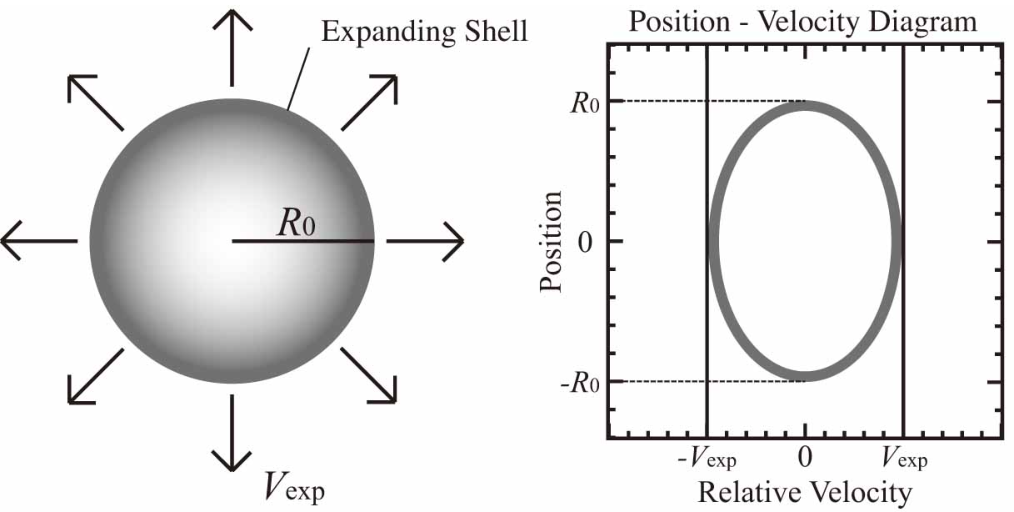}
\caption{Schematic image of a uniformly expanding shell and its velocity distribution in the position-velocity plane. Here we assume a radius of the shell $R_0$ and an expansion velocity $V_{\mathrm{exp}}$ of 9 pc and 10 km s$^{-1}$, respectively.}
\end{center}
\end{figure*}%

\begin{figure}
\begin{center}
\figurenum{B2}
\includegraphics[width=145mm]{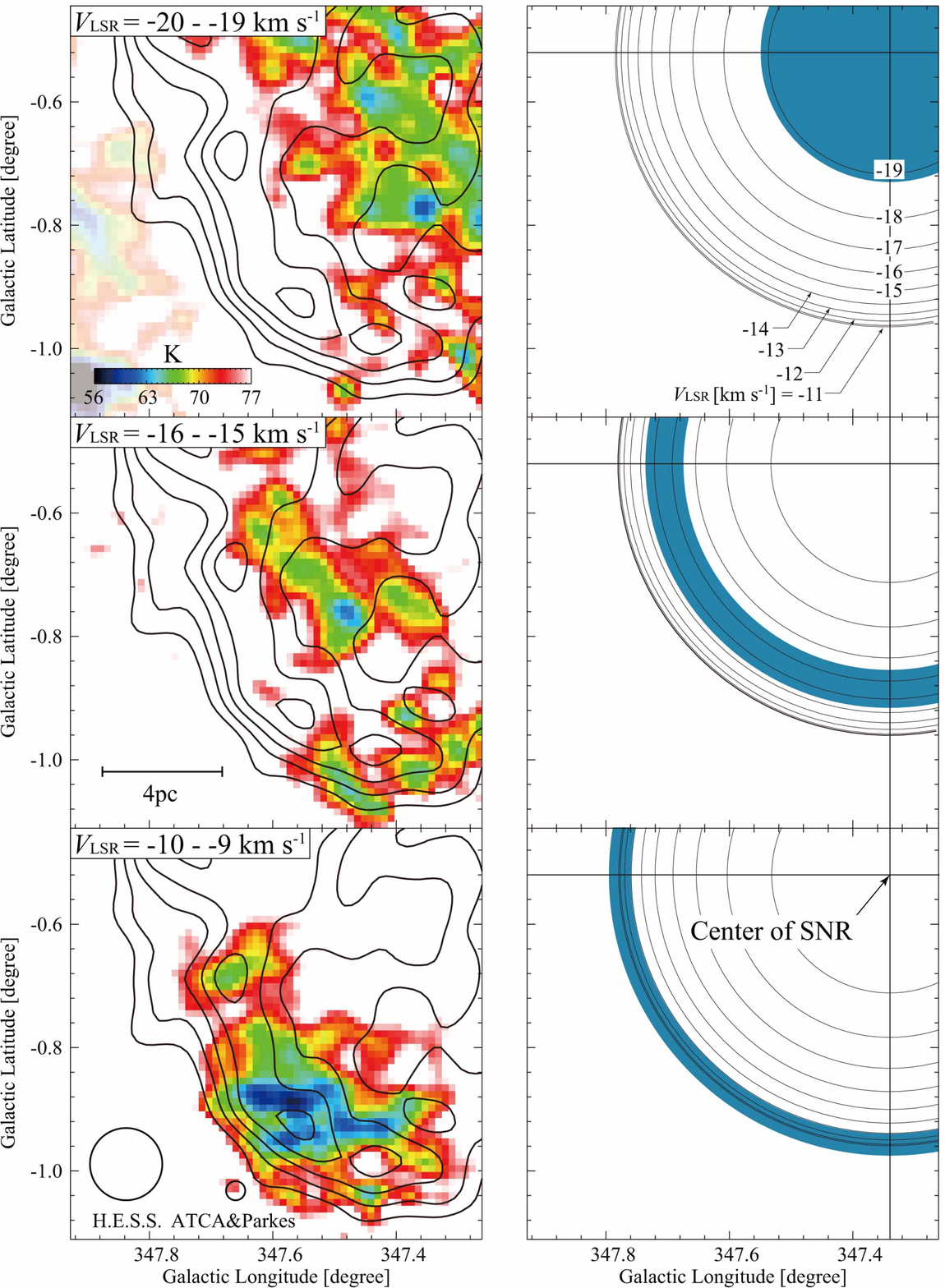}
\caption{(Left) Velocity channel distributions of \ion{H}{1} integrated intensity toward the SE cloud superposed on the TeV $\gamma$-ray contours. TeV $\gamma$-ray contours are plotted every 10 smoothed counts from 20 smoothed counts. The faded area in the upper panel is a component unrelated to the SNR. (Right) Model velocity distributions of an expanding shell shown in Figure B1. Iso-velocity lines are shown here, and blue areas show the corresponding velocity range shown in the left panels.}
\end{center}
\end{figure}%

\begin{figure}
\begin{center}
\figurenum{B3}
\includegraphics[width=158mm,clip]{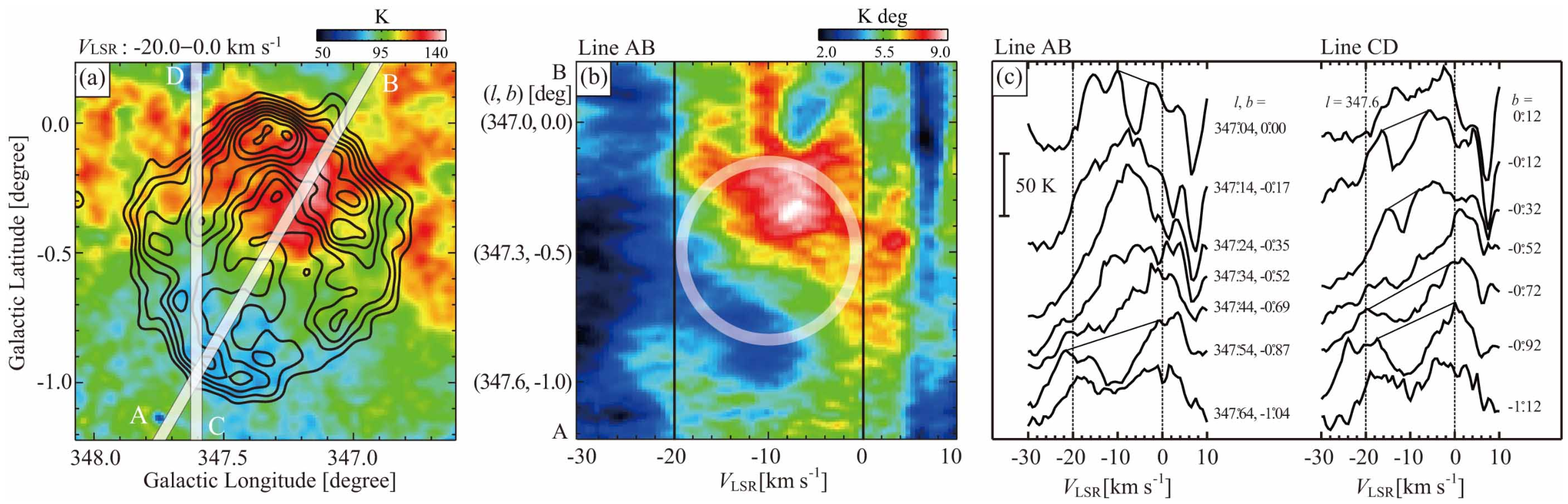}
\caption{(a) Averaged brightness temperature distribution of \ion{H}{1} in a velocity range from $-$20 km s$^{-1}$ to 0 km s$^{-1}$. Contours show the H.E.S.S. TeV $\gamma$-rays \citep{aharonian2007} and are plotted every 10 smoothed counts from 10 smoothed counts. The line AB is inclined by 60 degrees to the Galactic plane and the line CD passes the center of the SNR. (b) Position-velocity distribution of \ion{H}{1} along the line AB in Figure B3(a). The velocity resolution is smoothed to 1 km s$^{-1}$ and the integration interval is 200 arcsec. The white circle shows a schematic image of an expanding spherical shell (Figure B1). (c) \ion{H}{1} spectra along the lines AB and CD in Figure B3(a). Expected profiles of \ion{H}{1} self-absorption are shown by straight lines in the spectra with significant \ion{H}{1} dips.}
\end{center}
\end{figure}%

\begin{figure*}
\begin{center}
\figurenum{C1}
\includegraphics[width=158mm,clip]{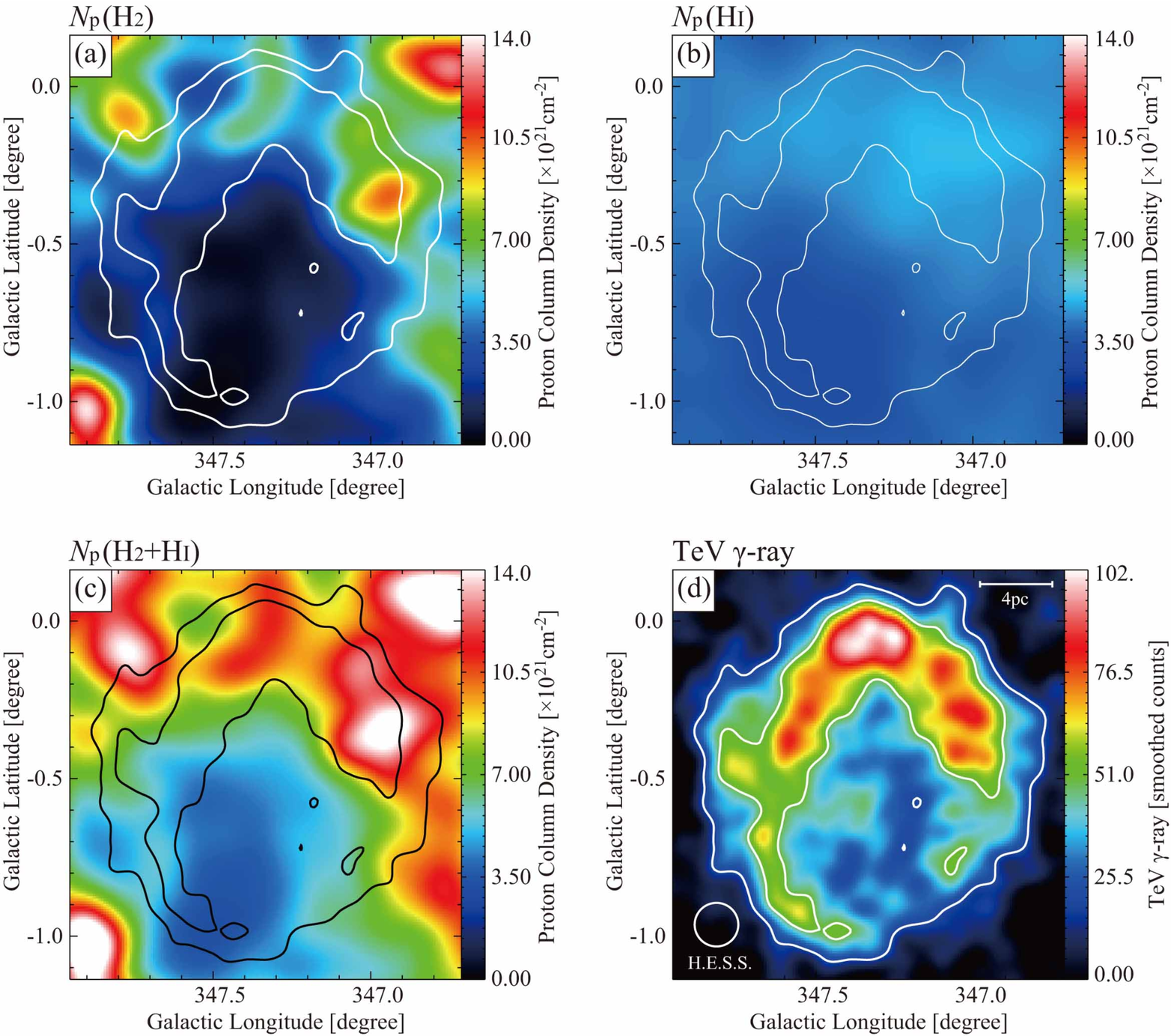}
\caption{(a) Distributions of column density of the ISM protons $N_{\mathrm{p}}$ estimated from $^{12}$CO($J$=1--0) $N_{\mathrm{p}}$(H$_2$), (b) \ion{H}{1} emission without correction for the self-absorption $N_{\mathrm{p}}$(\ion{H}{1}) and (c) sum of $N_{\mathrm{p}}$(H$_2$) and $N_{\mathrm{p}}$(\ion{H}{1}). Here we assume for reference that the \ion{H}{1} emission is optically thin and the \ion{H}{1} self-absorption is not taken into account. All datasets used here are smoothed to a HPBW of the TeV $\gamma$-ray distribution with a Gaussian function. (d) TeV $\gamma$-ray distribution. Contours are plotted every 50 smoothed counts from 20 smoothed counts.}
\end{center}
\end{figure*}%

\begin{figure*}
\begin{center}
\figurenum{C2}
\includegraphics[width=179mm]{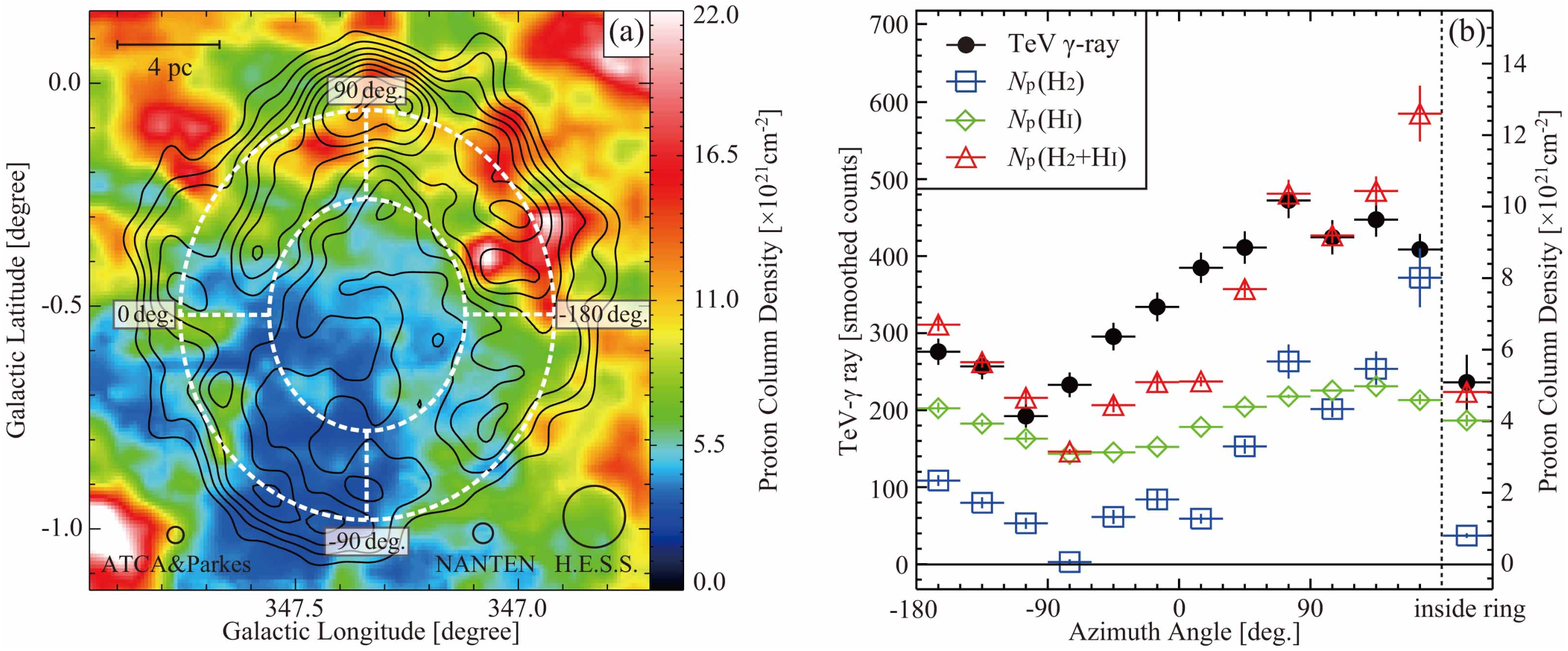}
\caption{(a) Distribution of column density of ISM protons $N_{\mathrm{p}}$(H$_2$+\ion{H}{1}) in a velocity range from $-$20 km s$^{-1}$ to 0 km s$^{-1}$, where the \ion{H}{1} is assumed to be optically thin and without self-absorption. Contours and two elliptical rings are the same as in Figure 8(a). (b) Azimuthal distributions of $N_{\mathrm{p}}$(H$_2$), $N_{\mathrm{p}}$(\ion{H}{1}), $N_{\mathrm{p}}$(H$_2$+\ion{H}{1}) and TeV $\gamma$-ray smoothed counts per beam in the two elliptical rings in Figure C2(a). The same plots inside of the inner ring are shown on the right side in Figure C2(b).}
\end{center}
\end{figure*}%

\begin{figure}
\begin{center}
\figurenum{C3}
\includegraphics[width=106mm]{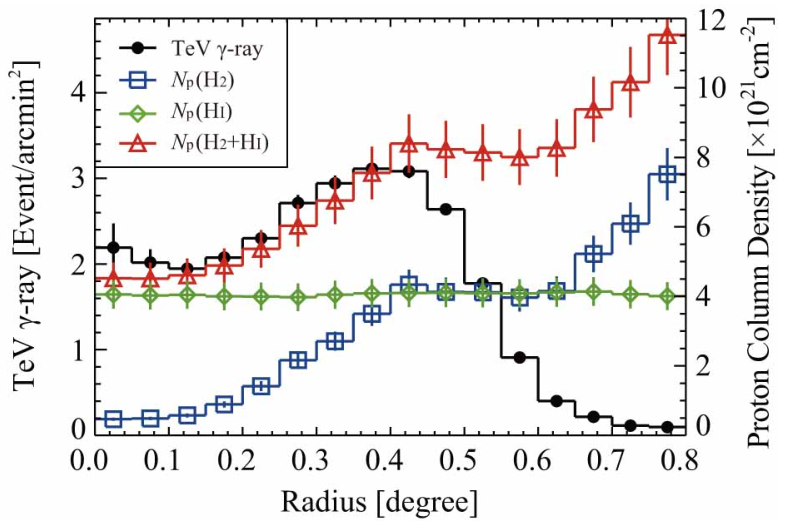}
\caption{Radial distributions of averaged values of TeV $\gamma$-rays, $N_{\mathrm{p}}$(H$_2$), $N_{\mathrm{p}}$(\ion{H}{1}) and $N_{\mathrm{p}}$(H$_2$+\ion{H}{1}), where the \ion{H}{1} is assumed to be optically thin as in Figure C1. $N_{\mathrm{p}}$(H$_2$) and $N_{\mathrm{p}}$(\ion{H}{1}) show column densities estimated from $^{12}$CO($J$=1--0) and \ion{H}{1}, respectively, and $N_{\mathrm{p}}$(H$_2$+\ion{H}{1}) shows the total ISM column density, the sum of $N_{\mathrm{p}}$(H$_2$) and $N_{\mathrm{p}}$(\ion{H}{1}).}
\end{center}
\end{figure}%

\end{document}